\documentclass{article}
\usepackage{emulateapj}
\usepackage{float,epsfig}

\usepackage{natbib}
\newcommand{\msun}{M$_{\odot}$}
\newcommand{\ergl}{ergs~s$^{-1}$}
\newcommand{\ergcms}{ergs~cm$^{-2}$~s$^{-1}$}
\newcommand{\DTF}{{$D_{25}$}}

\newcommand{\ros}{{\sl ROSAT}}
\newcommand{\xmm}{{\sl XMM-Newton}}
\newcommand{\cxo}{{\sl Chandra}}

\newcommand{\gmr}{$g-r$}
\newcommand{\umg}{$u-g$}
\newcommand{\rmi}{$r-i$}
\newcommand{\hii}{H\,{\sc ii}}
\newcommand{\hi}{H{\sc i}}
\newcommand{\ha}{H$\alpha$}

\slugcomment{Submitted to Astrophysical Journal}

\begin{document}

\title{Ultraluminous X-Ray Source Correlations with Star-Forming Regions}

\author{
Douglas~A.~Swartz\altaffilmark{1}
Allyn~F.~Tennant\altaffilmark{2}, and
Roberto~Soria\altaffilmark{3}}

\altaffiltext{1}{Universities Space Research Association, VP62,
    NASA Marshall Space Flight Center, Huntsville, AL, USA}
\altaffiltext{3}{Space Science Office, VP62,
    NASA Marshall Space Flight Center, Huntsville, AL, USA}
\altaffiltext{3}{Mullard Space Science Laboratory,
    University College London, Holmbury St. Mary, Surrey RH5 6NT, UK}

\begin{abstract}
Maps of low-inclination 
 nearby galaxies in Sloan Digitized Sky Survey \umg, \gmr\ 
 and \rmi\ colors are used 
 to determine whether Ultraluminous X-ray sources (ULXs) are predominantly 
 associated with star-forming regions of their host galaxies. 
An empirical selection criterion is derived from 
 colors of \hii\ regions in M\,81 and M\,101
 that differentiates between the young, blue stellar component 
 and the older disk and bulge population.
This criterion is applied to 
 a sample of 58 galaxies of Hubble type S0 and later
 and verified through an application of 
 Fisher's linear discriminant analysis.
It is found that 
 60\% (49\%) of ULXs in optically-bright environments are within regions  
 blueward of their host galaxy's \hii\ regions
 compared to only 27\% (0\%) of a control sample
 according to the empirical (Fisher) criterion.
This is an excess of 3$\sigma$ above the 32\% (27\%) expected if the ULXs
 were randomly distributed within their galactic hosts.
This indicates a ULX preference for  young,
 $\lesssim$10~Myr, OB associations.
However, none of the ULX environments have the morphology
 and optical brightness suggestive of a massive young super star cluster though several
 are in extended or crowded star-forming (blue) environments 
 that may contain clusters unresolved by Sloan imaging.
Ten of the 12 ULX candidates with estimated X-ray luminosities in excess of 
 3$\times$10$^{39}$~\ergl\ 
are equally divided among the group of ULX environments redward of \hii\ regions
 and the group of optically faint regions. 
This likely indicates that the brightest ULXs turn on at a time somewhat 
 later than typical of \hii\ regions; say 10$-$20~Myr after star formation
 has ended.
This would be consistent with the onset of an accretion phase as the 
 donor star ascends the giant branch if the donor is an $\lesssim$20~\msun\
 star. 
\end{abstract}


\section{Introduction}  

Ultraluminous X-ray sources (ULXs) are defined as the
most X-ray luminous off-nucleus point-like objects in nearby galaxies.
Their high luminosities ($L_{\rm X} \gtrsim 10^{39}$~\ergl\ in 
the few tenths to $\sim$10~keV range) imply high accretor masses; 
hence their interest as a possible evolutionary link 
between stellar-mass black holes ($M_{\rm BH} \sim 1-20$~\msun)
known throughout the Local Group and super-massive black holes 
 ($M_{\rm BH} \sim 10^6-10^9$~\msun) ubiquitous in the nuclei of galaxies.

Something beyond normal single-star evolution in the current epoch
 is required if ULXs are found to have masses much higher 
 than $\sim$100~\msun\  \citep{yungelson08, ohkubo06, heger03, heger02}.
Compact objects with masses $\gtrsim$100~\msun\ are often referred to 
 as intermediate-mass black holes, or IMBHs, to 
 emphasize this non-standard evolutionary path.

Star formation occurs on a wide range of mass scales, 
 even in the strongest starbursts,
 from the common small OB associations 
 up to the rare massive compact 
 super-star clusters (SSCs).
This hierarchy reflects 
 the mass distribution of their originating interstellar clouds
 and the different conditions present at the time of star formation.
For instance, the most massive clusters form only
 in high-pressure regions such as the high density 
 portions of dwarf galaxies and galactic nuclei 
 or the shocked regions of 
 interacting galaxies \citep{elmegreen97}.
Presumably, the most massive stars also form only in the most massive
 clusters \citep[e.g.,][]{elmegreen04, weidner06}.

Perhaps a similar hierarchical association between ULXs and their 
 local environments exists 
 with the lower-luminosity, presumably lower-mass, 
 ULXs affiliated with small unbound OB associations 
 and only the extremely luminous, potentially IMBH, objects 
 coincident with SSCs. 
Certainly, the observed correlation of ULXs with {\sl global} 
 star formation rate in spirals 
 \citep{swartz04, gilfanov04, liu06}
 implies that many bright ULXs are associated 
 with the short-lived population of massive stars.
ULXs are also more common and more luminous in starburst and 
 interacting/merging galaxies than in normal galaxies and
they occur at elevated rates in dwarf galaxies \citep{swartz08}.
Theoretically, the most 
 promising evolutionary path for IMBHs
 \citep[see the review by][for other possibilities]{vandermarel04}
 is merging of massive stars in the most dense young 
 star clusters \citep[e.g.,][]{pz04}.

What is needed is a systematic study of the {\sl local} environments of ULXs to 
 quantify any association ULXs may have with particular stellar systems and
 to quantify the properties (e.g., age and mass) of these systems. 
In this way additional constraints on the nature of ULXs can be inferred.

Constraints on the age of some ULXs and of their surrounding
stellar populations have been obtained from high-spatial-resolution
optical studies of individual objects using 
{\it Hubble} or ground-based 8-m class telescopes. 
These studies have found several examples of late O or 
 early B stars as either likely mass donors
 \citep{liu02, liu04, kuntz05, ptak06, terashima06, roberts08}
 or populating the nearby environment \citep[e.g.,][]{soria05, grise08}. 
In some cases, emission nebulae have been found associated with ULXs 
 \citep{pakull02, pakull06}.
These results suggest the more common ULX environment 
 is a small OB association rather than the compact core of a massive SSC. 
That is, if the hierarchical argument has any merit, the 
observational results to date argue against an IMBH association for ULXs.

However, these studies are feasible only for a few nearby ULXs
and are limited to those ULXs with luminous donors 
or at least to those located in OB associations. Hence,
they may not be representative of the average ULX population.
In fact, there are ULXs in elliptical galaxies. These  
likely have low-mass donors fainter than current detection limits
\citep[e.g.,][]{ptak06, roberts08}.

Here, a different approach is used
to quantify the strength of the association
between individual ULXs and local 
 stellar populations
that is applicable to larger samples of 
galaxies and out to larger distances 
where confusion prevents studies of individual stars
around a ULX.
Specifically, our goal is to determine what fraction of ULXs
are associated with \hii\ regions (currently star-forming regions,
or with an age $\la 10$ Myr since the end of star formation),
and what fraction is spatially coincident with young, massive
star clusters. To do so, we 
rely on area-averaged optical colors, instead of individual bright stars, 
 using the uniform, high-precision, 
 multi-band photometry provided by the Sloan Digitized Sky Survey (SDSS).
In particular, the SDSS $u$ band lies blueward of the $D(4000)$ break \citep{bruzual83}
 known to distinguish young from old stellar populations in SDSS data 
 \citep[e.g.,][]{kauffmann03} while $u-r$ or $u-g$ colors can distinguish
 early- from late-type galaxies in general 
 \citep{strateva01, baldry04}.

A description of the survey sample, 
 details of the reduction of SDSS images to color maps, 
 our definition of local environment, 
 and a summary of the X-ray data used in this work are presented in \S2.
An empirical criterion to distinguish star-forming regions 
 from non-star-forming regions  using SDSS colors is developed 
 in \S3  along with
 a complementary analysis using Fisher's linear discriminant analysis
 \citep{fisher36}.
Results of applying these criteria are given in \S4 and 
 the implications for ULXs is discussed in \S5.

\section{Sample Selection}

The galaxies included in this study are a subset of a complete
sample of 140 galaxies selected for X-ray study of their ULX populations.
The X-ray sample was chosen to form a volume-limited set of galaxies
 that are in both the set of 
 all galaxies within 15~Mpc contained in the Uppsala Galaxy Catalog 
 with photographic magnitude $m_p<14.5$~mag and the set of all galaxies listed in the 
 Infrared Astronomical Satellite ({\sl IRAS}) catalogs with a flux 
 $f_{\rm FIR}\ge10^{-13.3}$~\ergcms.
The UGC \citep{nilson73}
 contains all galaxies north of B1950 $\delta=-2^{\circ}30^{\prime}$
 in two complete samples: galaxies with angular diameters $>$1$^{\prime}$ on
 the first POSS blue prints and galaxies brighter than
 $m_p=14.5$ in the Zwicky Catalog of Galaxies and Clusters of Galaxies.
The IRAS fluxes were obtained from 60$\mu$m and 100$\mu$m 
 flux densities listed in either the 
 IRAS Cataloged Galaxies and Quasars \, \citep{fullmer89} or the
 IRAS Faint Source Catalog v2.0 \, \citep{moshir93} using the formula
 $f_{\rm FIR}/10^{-11}=3.25S_{60}+1.26S_{100}$ following \citet{ho97}
 where $S_{60}$ and $S_{100}$ are the total flux densities at 60 and 100~$\mu$m, respectively.
The {\sl IRAS} catalogs are complete to approximately 1.5~Jy for point-like sources.
The combined selection criteria favor nearby ($D<15$~Mpc) predominately  
 optically-bright galaxies with at least a modest amount of recent star formation 
 as inferred from their FIR luminosities. 
The sample is known to exclude smaller dwarf galaxies in the neighborhood 
\citep{swartz08}. 

Eighty-nine of the 140 galaxies comprising the X-ray sample are within
 the footprint of the
 6th data release of the Sloan Digitized Sky Survey \citep{dr6}.
Of these, only the 58 galaxies with inclination $i<65$\arcdeg\ make up
 the present study as
 it is too difficult to isolate local features in higher inclination galaxies.

Table~1 lists relevant properties of the galaxies included in this study.
Distances are adopted from the Nearby Galaxies Catalog \citep{tully88} 
 augmented by those computed using the heliocentric velocities listed in the 
 NASA/IPAC Extragalactic Database (NED) 
 assuming $H_o=73$~km~s$^{-1}$~Mpc$^{-1}$.
Galaxy celestial coordinates were taken from the NED.
Inclinations, optical shape (\DTF: major isophotal diameter measured at surface brightness level of 25 mag~s$^{-2}$ in $B$, position angle, and major-to-minor axis ratio) and morphological type are from 
 the Third Reference Catalogue of Bright Galaxies \citep{deVaucouleurs91}
Table~1 is composed of two sections: The 27 low-inclination galaxies hosting 
 one or more ULX and the 31 low-inclination galaxies without ULXs. 

\subsection{The SDSS Data}

All galaxies in the present study are extended objects by definition.
Therefore, SDSS pipeline products such as object lists are not of relevance here.
Instead, we begin with corrected imaging frames and use these to produce color maps
for our analysis. Details of this process are given in this subsection.

\subsubsection{Preparation of Color Maps}

Corrected frame images of the target galaxies in each of the 
$u$, $g$, $r$ and $i$ filters were retrieved from the 
SDSS archive\footnote{http://www.sdss.org/dr6/tutorials/retrieveFITS.html}. 
These are the imaging frames with flat field, bias, cosmic ray, and pixel defect corrections applied. Each covers an approximately 13.5\arcmin$\times$9.8\arcmin\
region of the sky. In several cases, one or more adjacent frames were
needed to cover a target galaxy; however, each image frame was processed separately.

There are often slight offsets among the 4 filter frames of a given field. The 
Montage(v3.0) image mosaic toolkit\footnote{http://montage.ipac.caltech.edu/}
module {\tt mProjectPP} was used to re-project the frames and module {\tt mSubimage}
was used to center and to crop a small border off of each frame. 
This ensured that the same sky coordinates were assigned to a given pixel
 in all 4 frames of a field.
Checks against known object positions were made to confirm that reprojection 
 did not compromise the astrometric accuracy of the frames.
This is important when registering ULX positions obtained from X-ray data 
 to the final color maps.

An estimate of the 
 sky level was then made by constructing an object mask that includes the 
 \DTF\ ellipse of the target galaxy, any bright objects elsewhere in the frame as
 determined by our source-finding software \citep{tennant06}, and
 a small border zone to include any artifacts introduced by the reprojection.
The sky level was then defined as 
 the average of the unmasked pixels
 in each frame (after properly accounting for the soft bias and converting pixel
data numbers to true counts using the gain parameters provided in the 
accompanying field information files).

Sky-subtracted corrected-frame data numbers were then 
 converted to calibrated magnitudes using the zero points, 
 extinction coefficients and airmass values also provided in the
 field information files.
Dereddening to correct for Galactic extinction was also applied using the 
  \hi\ maps of \citet{LAB05} 
 accessed through the FTOOL utility {\tt nh} and applying the extinction conversion
 for each SDSS filter as given in the early data release paper \citep{EDR}.

Finally, \umg, \gmr, and \rmi\ color maps were constructed by subtracting the appropriate 
 magnitude image in one spectral band from that in another.

\subsubsection{Defining Local Regions} \label{s:sdss_aperture}

Local regions are defined here to be the 100$\times$100~pc$^2$ area centered on
 the X-ray position of the ULX. 
This physical size was chosen to be roughly that of a moderate-sized \hii\ region. 
For comparison, the average diameter for the 248
\hii\ regions in M\,101 identified by \citet{scowen92} is 140~pc and the average 
half-light diameter of the 492 \hii\ regions in M\,81 compiled by
\citet{psk} is 56~pc. Similarly, OB stellar associations in Local Group galaxies
 are typically 80$-$100~pc diameter \citep[e.g.,][]{hunter95, gouliermis03}.

The colors of local regions are defined as the average of the colors of all pixels
 within the region. 
Note this is not equivalent to the color derived from the difference 
 of the average magnitudes in two spectral bands, i.e., it is not a flux-weighted
 but an area-weighted average color.
In this way, small bright objects are not overemphasized.
Our averaging method 
 results in colors in units of mag per 
 100$\times$100~pc$^2$ region although, 
 henceforth, color values in this work will be quoted simply in magnitudes with the
 fixed area implied.
Note this ``aperture photometry'' also differs from that of isolated point-like 
 objects for which a local background can be defined using 
 a nearby or surrounding source-free region. 
Here, neighboring regions contain similar sources of emission and 
 should not be considered a background. 
Thus, only the mean sky values are considered background in this work.

The 100$\times$100~pc$^2$ region, or resolution element, varies from 
 1.4\arcsec$\times$1.4\arcsec\
 (3.6$\times$3.6 pixels subtending 0.396~s~pixel$^{-1}$) for the most distant
 galaxy in our sample up to 9.8\arcsec$\times$9.8\arcsec\ 
 (24$\times$24 pixels) for the nearest target.
In order to maintain consistency when comparing local regions hosting ULXs
 to whole galaxies, the same procedure is applied to all 
 100$\times$100~pc$^2$ regions within the \DTF\ area of each galaxy
 (excluding those $<$3$\sigma$ above the sky level).
We report the resulting average galaxy colors which are area-weighted colors 
 defined as the sum over the colors of 
 all resolution elements within the \DTF\ ellipse of a galaxy,
 divided by the number of elements.
Average galaxy colors are listed in Table~1.
For reference, the
 number of resolution elements range from 334 for the 
 smallest galaxy (UGC~6850) to 53262 for the largest (M\,101) in our sample.

\subsection{The X-ray Data: ULX Candidates} \label{s:x_ulx}

The X-ray data for this study includes pointed observations obtained using \cxo,
\xmm, and \ros.
The source-finding and characterization program described in 
\citet{tennant06} was used to locate potential (point-like) ULXs. 
Source positions were compared to the NED
coordinates of the host galaxy nucleus to cull potential AGN from the ULX list.
These are the subject of another study \citep{zhang09}.
Similarly, comparison was made to catalogs of sources to identify and cull known
 background galaxies and foreground stars. 
Furthermore, visual inspection of the SDSS images was made to identify 
 potential interlopers not designated elsewhere.
(This final inspection resulted in the rejection of one foreground star; no known
 background sources were found coincident with ULX candidate positions.)
Applying the analytical form of the X-ray background log$N$$-$log$S$ distribution
 complied by \citet{moretti03} to the \DTF\ areas of each of the 58 galaxies 
 in our sample, we expect a total of 1.3 background sources bright enough
 to be classified as ULXs.
This number is small because 
 the typical angular size of our sample galaxies is modest and
 the galaxies are relatively nearby so that ULXs sample only the sparsely-populated
 high-flux portion of the background log$N$$-$log$S$ distribution.

Forty-seven ULX candidates were identified in 27 of the 58 low-inclination
galaxies in our sample. Their basic properties are listed in Table~2.
All ULX candidates were observed with \cxo's ACIS detector and these data are
 used to characterize their X-ray porperties.
Some of the (\cxo) observations are deep enough that the 
 luminosities of ULX candidates can be accurately
 estimated using spectral fitting 
\citep[taken from the literature or performed here following the methods of][]{swartz04}.
Other observations are short exposures that
 accumulated only 20$-$50 events per ULX. 
For these short \cxo\ exposures, a straight counts-to-flux conversion is applied to
 estimate source luminosities as indicated in Table~2.
The conversion assumes a power-law source spectrum
 of photon index 1.8 \citep{swartz04}
 absorbed by the Galactic hydrogen column density along the 
 line of site \citep[from][]{LAB05}. 
The conversion is roughly 
 $f_{\rm X} \sim 10^{-11} \dot{N}$~\ergcms\ where $f_{\rm X}$ is the flux  
 and $\dot{N}$ is the observed count rate of events in the 0.5$-$8.0~keV band.
The exact conversion depends on the line-of-sight absorption. 
This counts-to-flux conversion was also applied to the deep \cxo\ observations 
 to test the validity of the method; 
the simple estimates agree with published luminosities, where available, to
 within a factor of four which is adequate for our present purposes.

\section{Star-formation Color Criteria} 

Two methods are employed to 
differentiate between the young stellar component 
 and the older disk and bulge population in galaxies based on their SDSS colors.
The first (\S\ref{s:ECC}) is a simple empirical method that selects 
 regions bluer in all three colors than \hii\ regions.
The second (\S\ref{s:LDA}) is a more rigorous analytical method 
 that finds the linear combination of colors which best separate star-forming 
 from non-star-forming regions.
The two methods give very similar results.

\subsection{Empirical Star-Formation Criteria} \label{s:ECC}

Here we establish,
empirically, the SDSS
 colors of regions known by independent means to be sites of recent 
 star formation. 
For nearby galaxies, an appropriate and easily-accessible
 recent star formation indicator 
is \ha\
 emission from \hii\ regions.
This \ha\ emission comes from recombination of gas photoionized by
massive O and early B stars and therefore indicates
ages less than $\sim$10~Myr. 
These stars themselves emit strongly in blue light.
We therefore expect \hii\ regions to be bluer than non-star-forming regions
 in the broad SDSS filters.
There are, however, two factors that  could lead to redder colors;
dust obscuration 
and the underlying stellar continuum. 
Thus, for instance,
 a low-luminosity \hii\ region located within the bright bulge of a galaxy
 can appear more red than a similar region located in the disk.

\begin{figure*}
\begin{center}
\includegraphics[angle=-90,width=0.28\columnwidth]{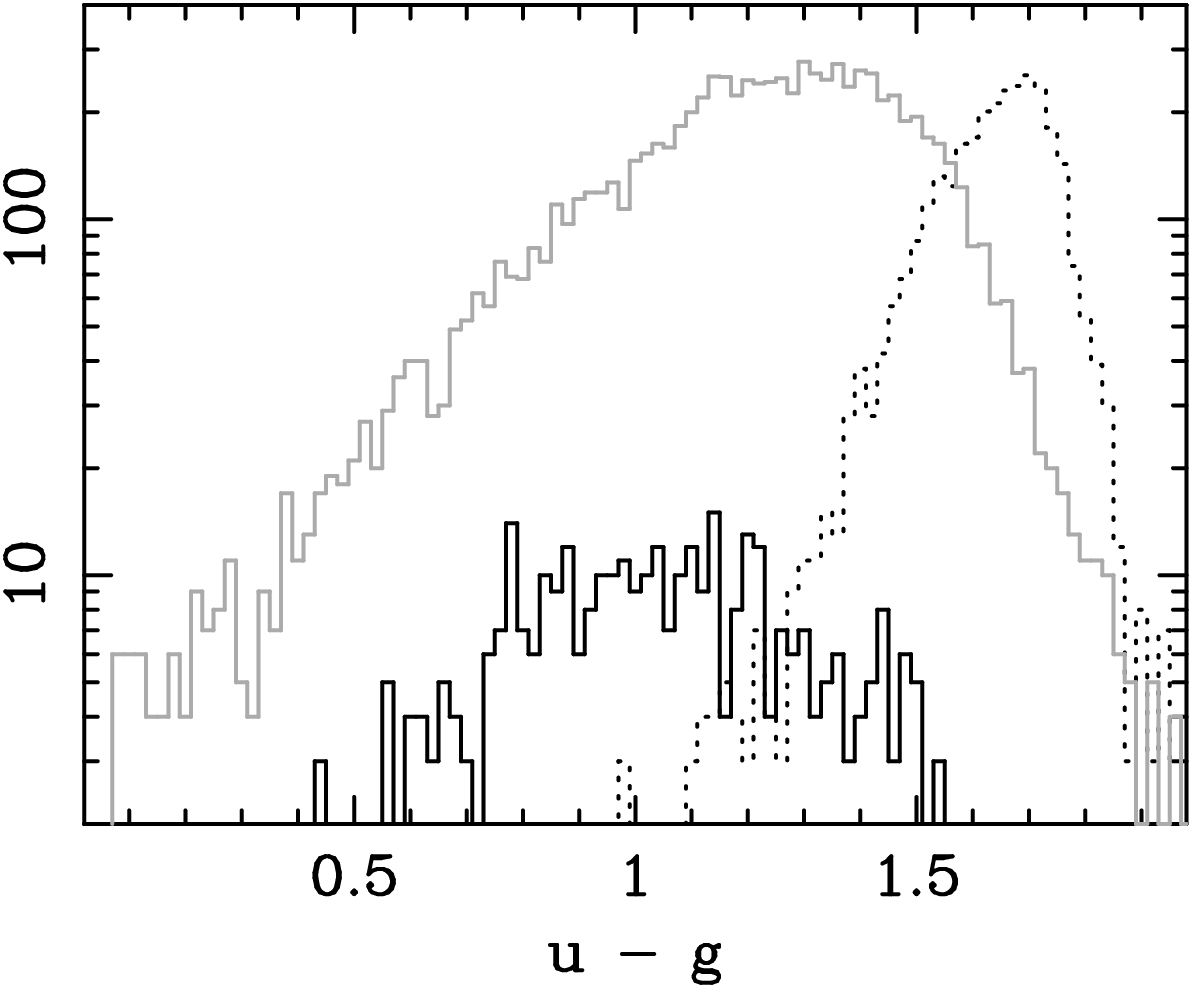}
\hspace{5pt}
\includegraphics[angle=-90,width=0.28\columnwidth]{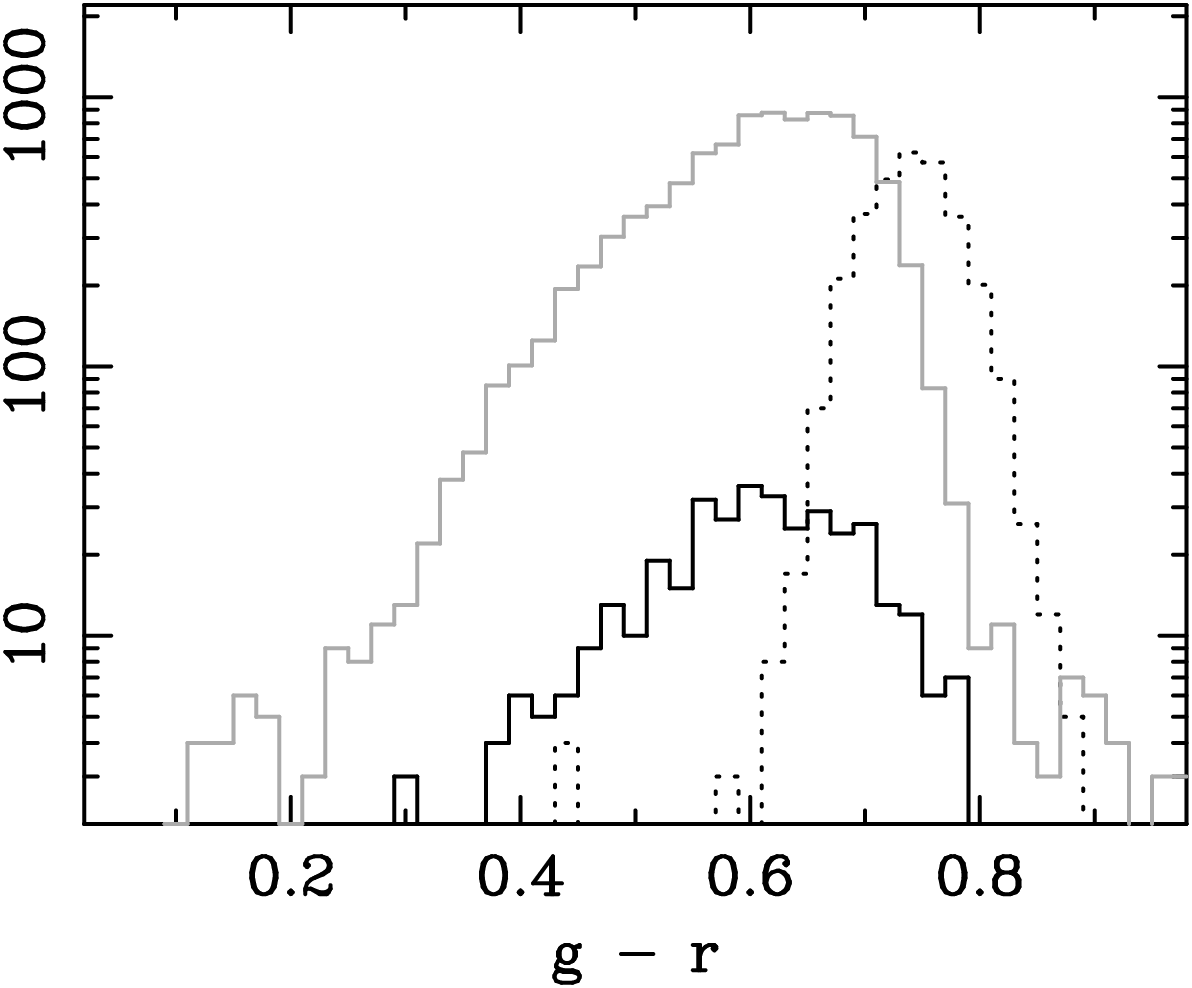}
\hspace{5pt}
\includegraphics[angle=-90,width=0.28\columnwidth]{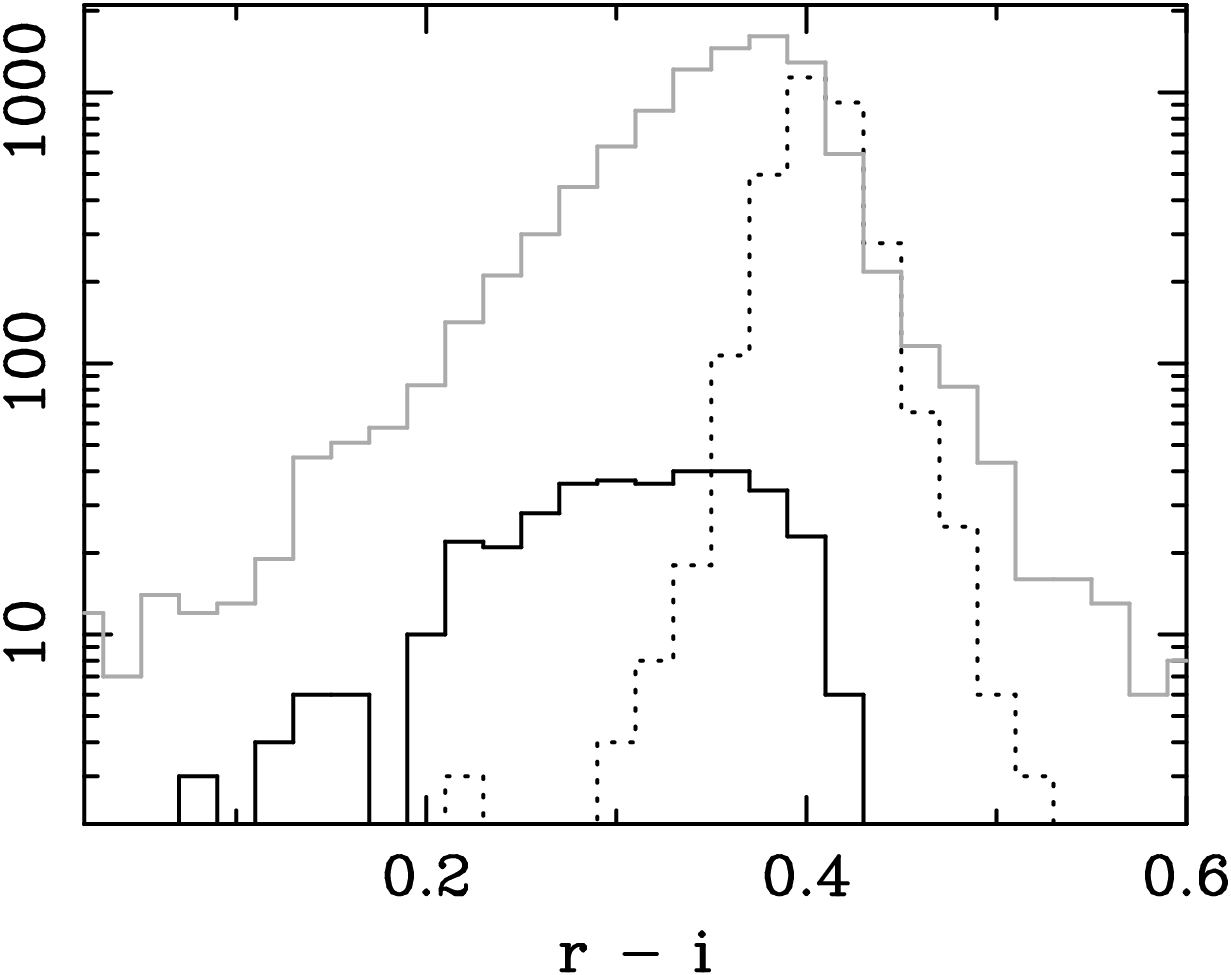}
\hspace{15pt}
\includegraphics[angle=-90,width=0.28\columnwidth]{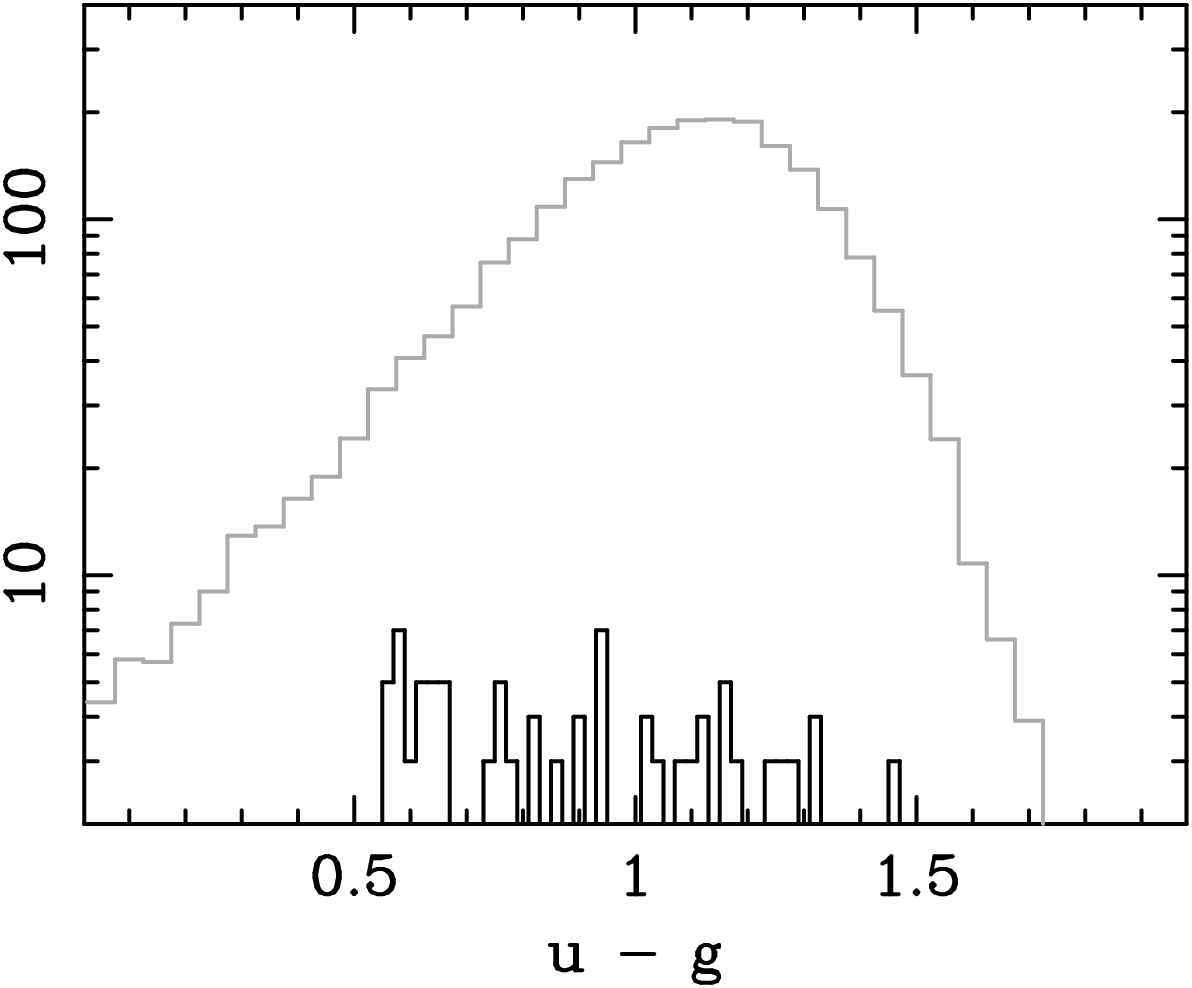}
\hspace{5pt}
\includegraphics[angle=-90,width=0.28\columnwidth]{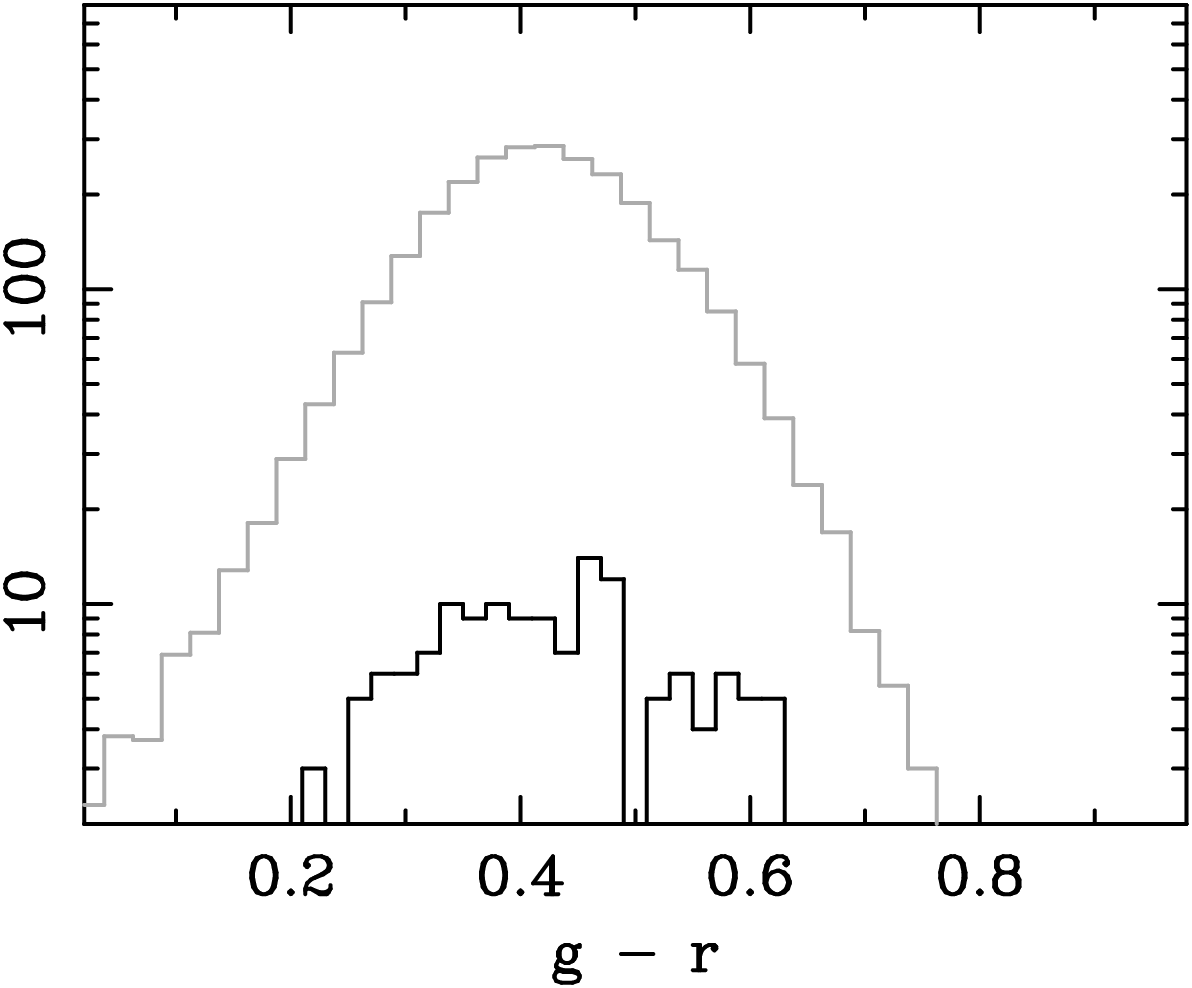}
\hspace{5pt}
\includegraphics[angle=-90,width=0.28\columnwidth]{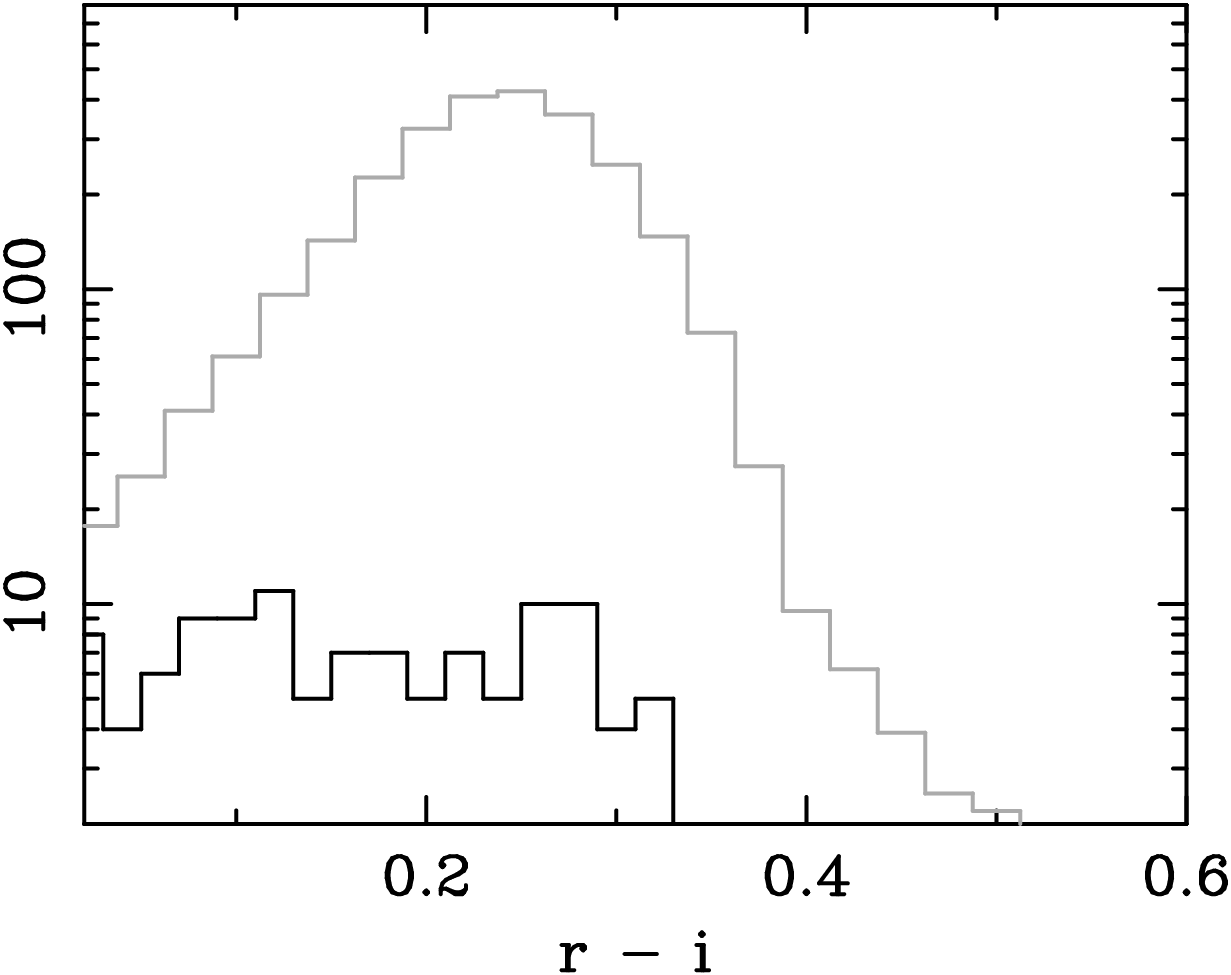}
\figcaption{Color distributions in the bulge, disk, and \hii\ regions of M\,81 
({\sl top}) and the disk and \hii\ regions of M\,101 ({\sl bottom}).
{\sl Left-to-right:} \umg, \gmr, and \rmi\ colors of all 100$\times$100~pc$^2$
regions in the color map; {\sl Heavy line:} \hii\ regions, {\sl light line:} disk, 
{\sl dotted line:} bulge. 
\label{f:m81_ih}}
\end{center}
\end{figure*}

To develop an emprical star-formation criteria, 
 we examine two well-studied large galaxies in our sample;
 the early-type SA(s)ab spiral M\,81 (NGC~3031) and the late-type SAB(rs)cd galaxy 
 M\,101 (NGC~5457).
The top three panels of Figure~\ref{f:m81_ih} show the distribution of 
 colors in the central bulge, disk, and cataloged \hii\ regions 
 \citep[whose locations and angular sizes are taken from][]{psk} 
 of M\,81 in each of the three colors \umg, \gmr, and \rmi.
The average of the colors 
 of the \hii\ regions and of the  colors of the galaxy as a whole are
 listed in Table~3 (where errors denote one standard deviation).
As expected, the \hii\ regions are bluer than the bulge and the disk
but there is considerable scatter which is due in part to obscuration by dust.

We performed a similar analysis for M\,101 (bottom panels of Figure~\ref{f:m81_ih}
and Table~3). We find, first of all, that 
 the mean colors of the galaxy are bluer than those of M\,81
 due to the large contribution to the average colors from the (red) bulge of M\,81 
 in contrast to the (blue) disk-dominated late-type galaxy M\,101.
The means of the colors of \hii\ regions 
 \citep[sizes and locations from the tabulation of][]{scowen92} are also
bluer than those of M\,81 (Table~3).
This indicates that the SDSS colors of star-forming regions are, indeed, 
 determined in part
 by the underlying stellar light contribution and will vary from 
 galaxy to galaxy.

To summarize, we may formulate our criterion to distinguish star-forming regions
based on the following empirical evidence:
(1) Star-forming regions are bluer in all three SDSS colors, in general, 
 than are non-star-forming regions in the same galaxy. 
(2) Allowance must be made for reddening due to dust.
(3) Allowance must be made for differences among galaxies due to differences
 in their underlying stellar continua.
(4) Allowance must also be made for low $S/N$ regions for which colors are 
 poorly constrained.

Thus, we adopt the following criterion 
 to distinguish star-forming from 
 non-star-forming regions.
The mean colors of the M\,81 \hii\ regions define the baseline. 
These values are scaled upward by 1$\sigma$ to allow for dust reddening. 
For each host galaxy, these values are scaled further
 by the difference of the mean colors of the host galaxy to that of M\,81 to allow
 for variations in the underlying stellar continua.
Symbolically, a region in the $i^{th}$ galaxy is flagged as a star-forming region
 if, for all three colors $C_{j,i}$, $j=1,3$, of the region
\begin{equation} \label{e:color_def}
C_{j,i}<(h_j+\delta h_j) - ( \langle C_{j,\rm M81} \rangle - \langle C_{j,i} \rangle )
\end{equation}
where $h_j$ is the mean color
and $\delta h_j$ is the standard deviation for \hii\ regions in M\,81,
 $\langle C_{j,\rm M81} \rangle$ is the average color of all
 100$\times$100~pc$^2$ regions in M\,81,
 and $\langle C_{j,i} \rangle$ is the average color of regions
 in the $i^{\rm th}$ galaxy.
In addition,
 regions fainter than 3$\sigma$ above the mean sky level in any of the 
 four filters are rejected as low $S/N$ regions under criterion (4) above.

Applying Equation~\ref{e:color_def} to M\,101 results in color criteria of
 $C_{j, \rm M101}<1.07$, 0.50, and 0.27 mag for \umg, \gmr, and \rmi, respectively. 
These are close to the average colors (plus 1$\sigma$) 
 of the known \hii\ regions of this galaxy (Table~3).
Table~1 (see \S~\ref{s:sdss_aperture}) includes the mean galaxy colors,
 $\langle C_{j,i} \rangle$, 
 and the fraction, $f_{\rm HII}$, of 100$\times$100~pc$^2$ regions classified as 
 bluer than typical \hii\ regions
 according to Equation~1 for each galaxy in the sample.

\subsection{Fisher's Linear Discriminant} \label{s:LDA}

The empirical criterion, Equation~1, 
  treats the 3 SDSS colors as independent quantities.
The possibility exists that some other combination of colors or magnitudes
 may better distinguish star-forming regions from other regions.
To test this hypothesis, 
 we used the ``Toolkit for Multivariate Data Analysis with 
 ROOT'' or TMVA\footnote{http://tmva.sourceforge.net/}, an add-on package
 to the script interpreter ROOT available from 
 CERN\footnote{http://root.cern.ch/}.
This Toolkit provides several methods of
 multivariate analysis to separate classes of objects.
In applying the evaluation and test phases of the Toolkit,
 we found that linear discriminant analysis,
 in the form first described by \citet{fisher36}, performed 
 very well and we adopted it here.
The Fisher 
 algorithm
 finds the linear combination of attributes 
 that best separates two or more classes of objects.
In our case, the attributes are the SDSS magnitudes and colors
 and we wish to distinguish the class of star-forming regions 
 from the class of non-star-forming regions.
The algorithm maximizes the ratio of the distance between the classes
 to the variance within the classes thereby guaranteeing maximal separability.
The output is a set of linear coefficients, $a_i$ for the
 input attributes, $x_i$ that gives the Fisher index, 
 $F=a_0+a_1x_1+a_2 x_2+ \cdots$, such that, in our case, $F>0$ would denote
 a star-forming region.

We began with the known class of \hii\ regions in M\,81
 as the training set ``source'' and all 100$\times$100~pc$^2$ regions in M\,81
 (that are $>3\sigma$ above the sky level) as the training set ``background''.
Inspection of the resulting Fisher index indicated that SDSS colors
 were more important attributes than magnitudes
 for separating star-forming from non-star-forming regions; which makes
 sense physically.
Furthermore, there appeared to be little benefit gained by 
 including all three colors with \gmr\ being the least important.
This naturally suggested using just the two attributes, \umg\ and \rmi.
Retraining using only these two attributes results in a Fisher index 
\begin{equation}
F = 0.374 - 0.179 (u-g) - 0.491 (r-i).
\end{equation}

Equation~2
 did a very good job of separating young star forming regions in M\,81
 and provides a criterion comparable to the empirical method developed
 in \S\ref{s:ECC}.

\begin{figure*}
\begin{center}
\includegraphics[angle=-90,width=0.45\columnwidth]{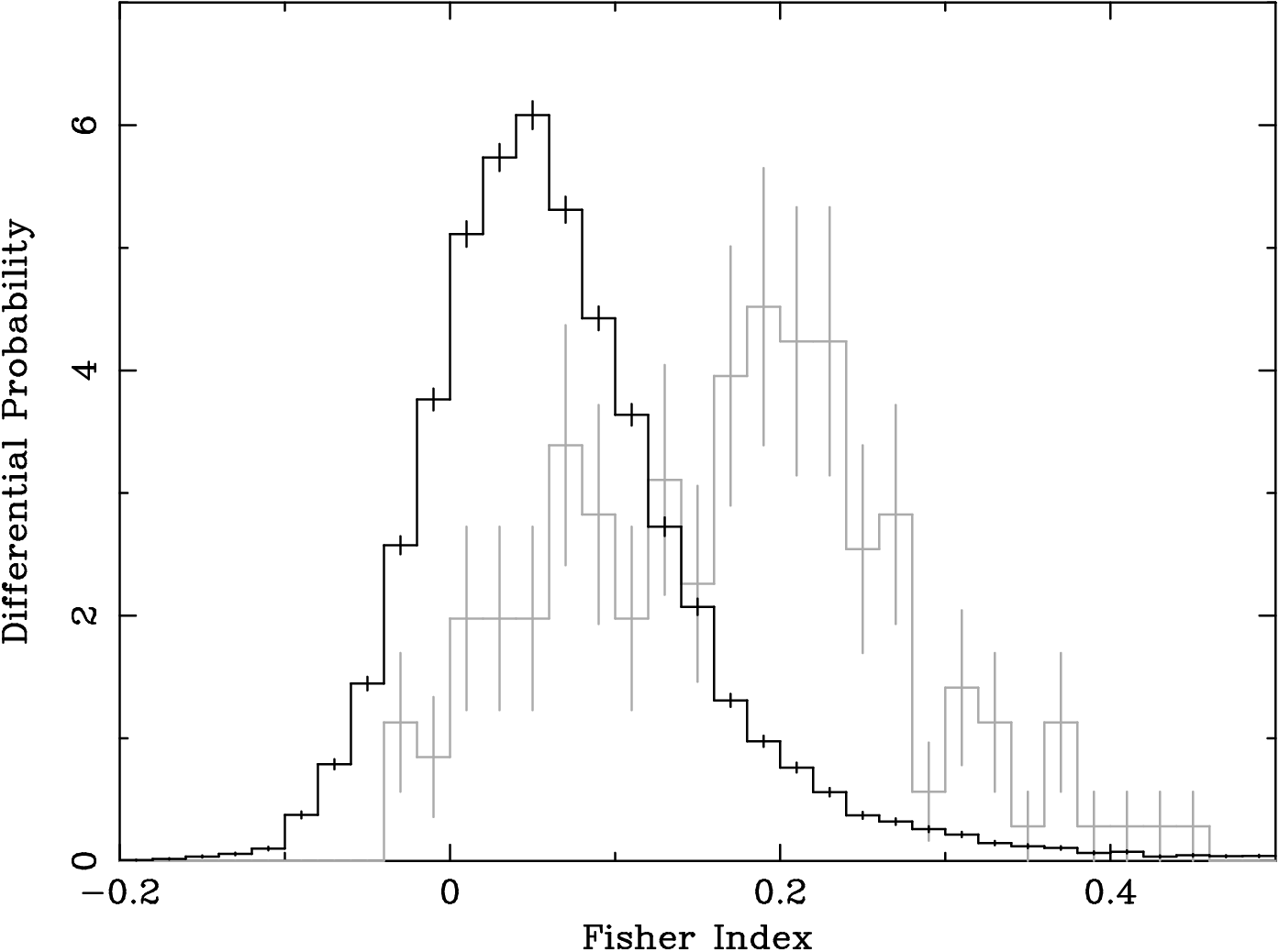}
\hspace{15pt}
\includegraphics[angle=-90,width=0.45\columnwidth]{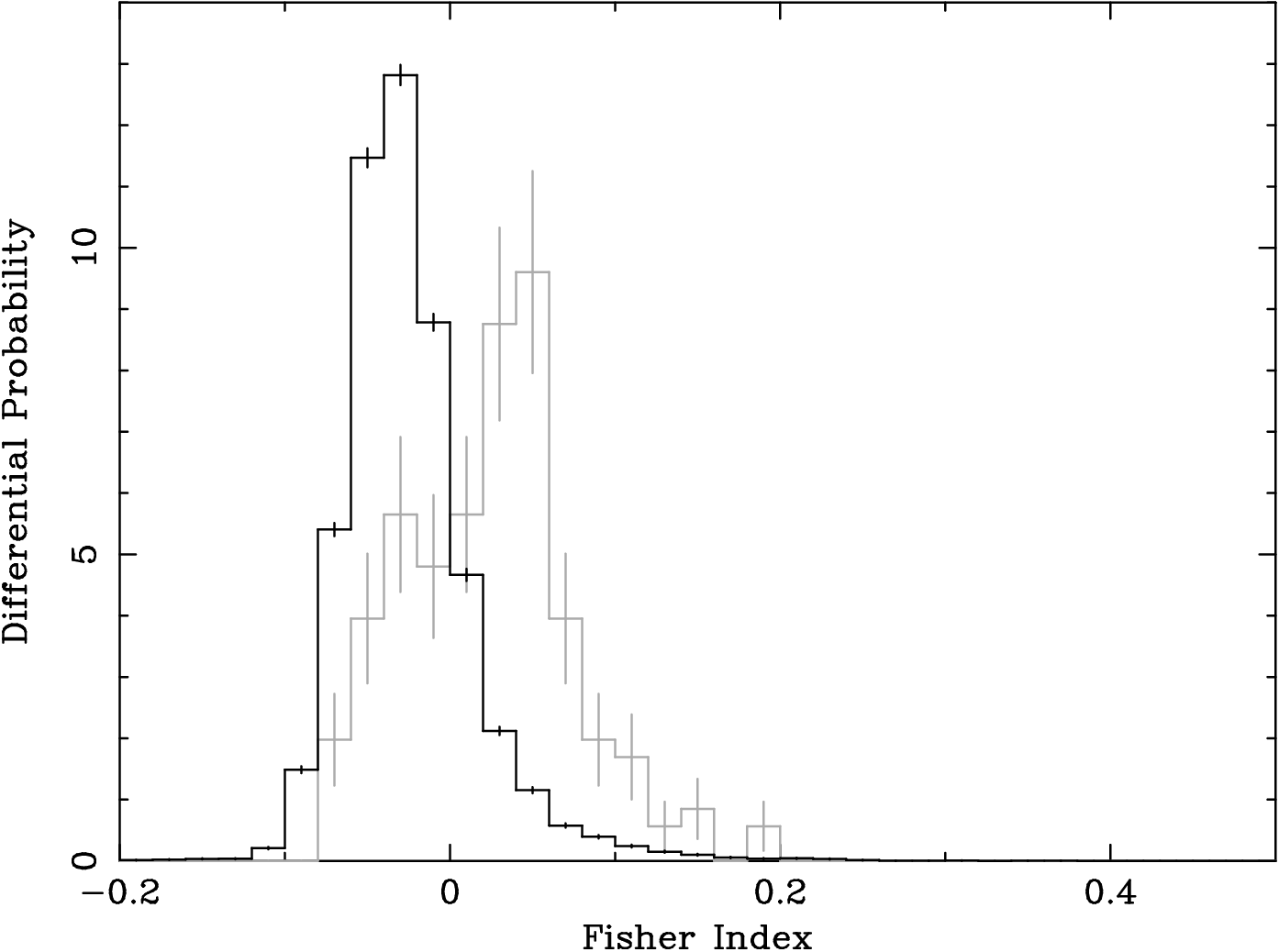}
\figcaption{
{\sl Left:} 
Image of the central field of M\,81 
with all regions classified as star-forming according to the 
Fisher linear discriminant ($F>0$, Equation~2) highlighted.
This result can be be compared to the empirical method displayed in the
rightmost panel of Figure~\ref{f:m81_ih}.
{\sl Middle:} 
Distribution of all 100$\times$100~pc$^2$ regions of M\,101 in the 
Fisher index (heavy line) using the M\,81 \hii\ regions as the
training set (Equation~2). The light line traces the histogram of
the known \hii\ regions in M\,101.
The area under each histogram is normalized to unity. 
{\sl Right:} 
Same as middle panel but with the Fisher index trained using the \hii\
 regions of M\,101. Note the fiducial $F=0$ separator applies here 
but that $F\approx 0.1$ would be more appropriate for the result 
shown in the middle panel. See text for further discussion.
\label{f:fm81_m101}}
\end{center}
\end{figure*}

Applying Equation~2 to M\,101 classifies most of the galaxy as star-forming
 because, as before, the mean colors of the 
 early-type galaxy M\,81 are redder than those of M\,101.
This is shown graphically in Figure~\ref{f:fm81_m101}.
The left panel shows the distribution of M\,101 regions using the 
 \hii\ regions of M\,81 as the training set (Equation~2). 
The right panel shows the distribution resulting after retraining using
 the \hii\ regions of M\,101 itself. 
Both clearly separate the star-forming \hii\ regions from non-star-forming
 regions although the distributions differ in several respects.
The most important difference is
 the overall shift or offset from the fiducial $F=0$ separation criterion
 to $F\approx 0.1$ when using the M\,81 training set.  
Noting that the average Fisher index for M\,81
 (computed using the average colors
 tabulated in Table~1 to evaluate Equation~2) is $-0.036$
 and that for M\,101 is $+0.074$ suggests that the Fisher index
 can be easily compensated for differences in the underlying 
 galaxy colors by applying an offset 
 $\Delta_i=(\langle F_{i}\rangle - \langle F_{\rm M81}\rangle)$
 such that a region in the $i^{th}$ galaxy is flagged as a star-forming region 
 if $F>\Delta_i$. 
Here, $\langle F_{i} \rangle$ is the value of $F$
 from Equation~2 evaluated using the average \umg\ and \rmi\ colors of the
 $i^{th}$ galaxy and $\langle F_{\rm M81} \rangle$ is the corresponding
 value for M\,81; $\langle F_{\rm M81} \rangle = -0.036$. 
This results in $\Delta=0.109$ for M\,101. 
This is wholly equivalent to the scaling by the mean galaxy colors 
 that is part of the empirical criterion (item 3) leading to Equation~1.

A comparison of the two methods 
 is given by the fraction of \hii\ regions 
 actually selected as star-forming. 
These are listed in Table~3 for M\,81 and M\,101. 

\begin{figure*}
\begin{center}
\includegraphics[angle=-90,width=0.90\columnwidth]{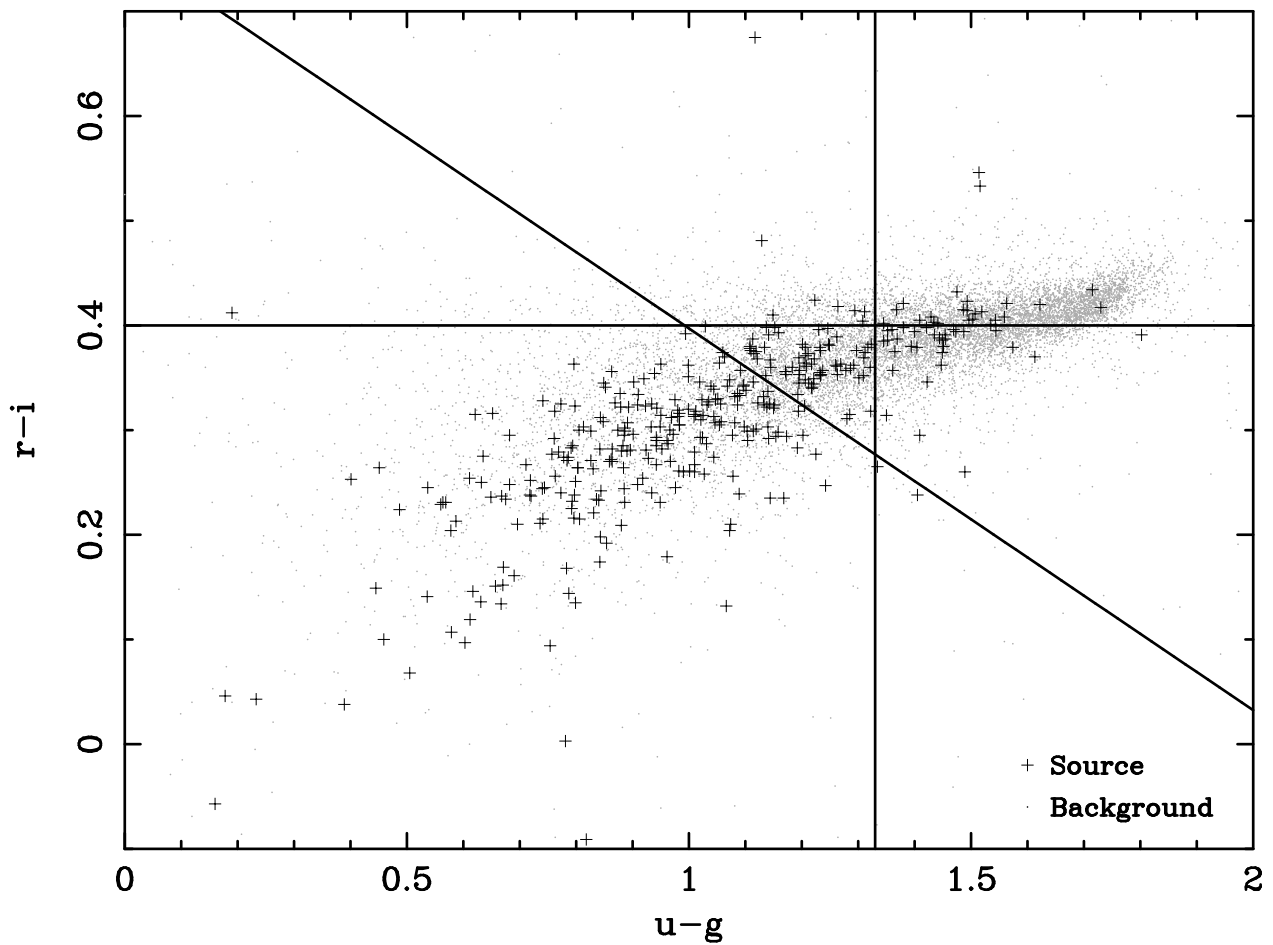}
\figcaption{Color-color diagram of the 100$\times$100~pc$^2$ regions in 
 M\,81. Known \hii\ regions \citep{psk} are indicated by crosses.
The horizontal and vertical lines are the empirical delineations 
(Equation~1) and the diagonal line is the Fisher index, $F\equiv0$
(Equation~2) separating blue star-forming regions below and to the left of these
 lines and red bulge and disk regions above and to the right.
\label{f:colcol}}
\end{center}
\end{figure*}

Finally, a direct comparison of the two methods is illustrated in
 Figure~\ref{f:colcol}. 
It is a color-color diagram for the M\,81 training data.
Each point represents the \rmi\ vs. \gmr\ colors of a 
 100$\times$100~pc$^2$ region in the galaxy with the
 known \hii\ regions \citep[from][]{psk} highlighted.
The diagonal line traces $F=0$ from Equation~2 and the 
  horizontal and vertical lines represent two of the three empirical
  cuts from Equation~1 and Table~3.
Regions to the lower left in the figure are designated blue star-forming
 regions according to these critiria.
Note that the colors of M\,81 regions at the red portion of the 
 diagram, $(u-g,r-i)\approx(1.8,0.5)$, correspond to those 
 of K0 main sequence stars, as expected, whereas those in the blue
 portion, say $(u-g,r-i)\approx(1.0,0.3)$, do not correspond to colors
 of a single stellar population; $u-g=1.0$ corresponds to B8 to A0 stars
 whereas $r-i=0.3$ is the color of G0 to G5 stars. 
This is further evidence of the contribution to young star-forming 
 regions of the older underlying disk and bulge populations. 

\section{Results} \label{s:applied}

The two criteria developed in the previous section provide 
 similar results. 
We emphasize  results from the empirical method, \S\ref{s:ECC},
 in the following
 but also quote values from the Fisher linear discriminant when
 appropriate.

\subsection{Quantitative Considerations}

We can divide the 100$\times$100~pc$^2$ regions surrounding the 
 47 ULX candidates in our sample into three non-overlapping groups:
Those rejected as being too faint 
 for quantitative analysis of their colors;
those meeting the star-forming criterion; and
those that do not meet the star-forming criterion. 
For brevity, these groups will be designated as $R$ for rejected, $S$ for star-forming,
 and $N$ for non-star-forming, respectively; they are listed in Table~2.
There are 12 ULX candidates in $R$ regions.
Using the empirical criterion, there are 21 ULX candidates in $S$ regions
 and 14 in $N$ regions.
The results are similar for the Fisher criterion except that
6 regions identified as $S$ by Equation~1 are classified 
 as non-starforming by the Fisher criterion 
 and 2 regions identified as $N$ are classified as star-forming. 
Inspection of the SDSS 3-filter 
 images\footnote{available on-line from 
          http://cas.sdss.org/astro/en/tools/chart/chart.asp}
 did not suggest a reason for this discrepancy. 
However, we note that the two methods agree well in the statistical sense.

As a simple check against the possibility of introducing biases into our analysis
 due to large-scale structures within ULX host galaxies, we queried 
 a control sample of regions in each galaxy and analyzed them in a manner
 completely analogous to that of regions surrounding the ULX candidates.
This sample is defined as the 100$\times$100~pc$^2$ regions diametrically opposed to
 the ULX regions in each host galaxy.
In this way, at least for well-ordered spirals, the control group samples regions similar
 to those of the ULXs on large scales but not on the small scales of individual 
 star-forming regions. 
Using the same group definitions as given above for the ULX population, the 
 control sample is distributed as 21(21) group $R$ objects, 7(0) group $S$ objects, 
 and 19(26) group $N$ objects using the empirical (Fisher) criterion.
This indicates the ULXs are preferentially located in star-forming regions and that 
 these regions are more luminous than their surroundings. 

We can also estimate the probability of a random sampling of star-forming regions.
We applied the empirical 
 star-formation criteria to all 100$\times$100~pc$^2$ regions in the
 27 galaxies containing ULXs.
This ranks 31.8\% of all regions bright enough for analysis as star-forming. 
Thus if ULXs occurred randomly, we would expect only 11
 of the 35 to be classified as star forming regions.
The chance of finding 21 when only 11 are expected
 is a 3.1$\sigma$ deviation.
On the other hand,
 finding that 7 of 26 control sample regions, or 27\%, are star-forming 
 is wholly consistent with a random distribution of control sample regions.
The more restrictive 
 Fisher criterion classifies only 23.5\% of regions as star-forming.
The chance of finding 17 ULXs in star-forming regions 
 when only 8 are expected is a 3.1$\sigma$ deviation
while finding 0 of the control sample in star-forming regions
 when 5 are expected is 2.3$\sigma$ low. 

\subsubsection{Correlations with Star-forming Regions}

Fully 60\% of those ULX regions 
 with sufficient signal ($21/35$) are in star-forming regions.
By definition, these regions are as blue or bluer than typical \hii\ regions 
 of their host galaxies and 
 therefore are likely of an age also typical of \hii\ regions which is $\lesssim$10~Myr;
 the characteristic lifetime of the least-massive LyC-producing stars, i.e., late~O to
early~B stars of about 15 to 20~\msun.

\subsubsection{Correlations with Host Galaxy Morphology}
Ten of the 27 galaxies hosting ULX candidates in our sample are early-type 
 galaxies (S0 through Sbc). 
There is no preference for objects in any of the 3 groups, $R$, $S$, or $N$, to be
 associated with a particular galaxy morphological type.

\subsubsection{Correlations with ULX Luminosity}
The average X-ray luminosity of group $S$ objects is only 1.7$\pm$0.7
 in units of 10$^{39}$~\ergl. The average of group $N$ is 3.5$\pm$3.6 and of group $R$
 is 3.9$\pm$3.8. 
These numbers reflect the fact that most (74\%) 
 of the ULXs in the current sample have observed luminosities estimated to be 
 $<$3$\times$10$^{39}$~\ergl\ and that 10 of those 12 more luminous than this value
 are equally divided among groups $N$ and $R$. 
Thus, the average luminosities in these groups are high
 while the scatter in their luminosities is large.

\subsubsection{Comparison to Mean Galaxy Colors} \label{s:meancolors}

\begin{figure*}
\begin{center}
\includegraphics[angle=-90,width=0.31\columnwidth]{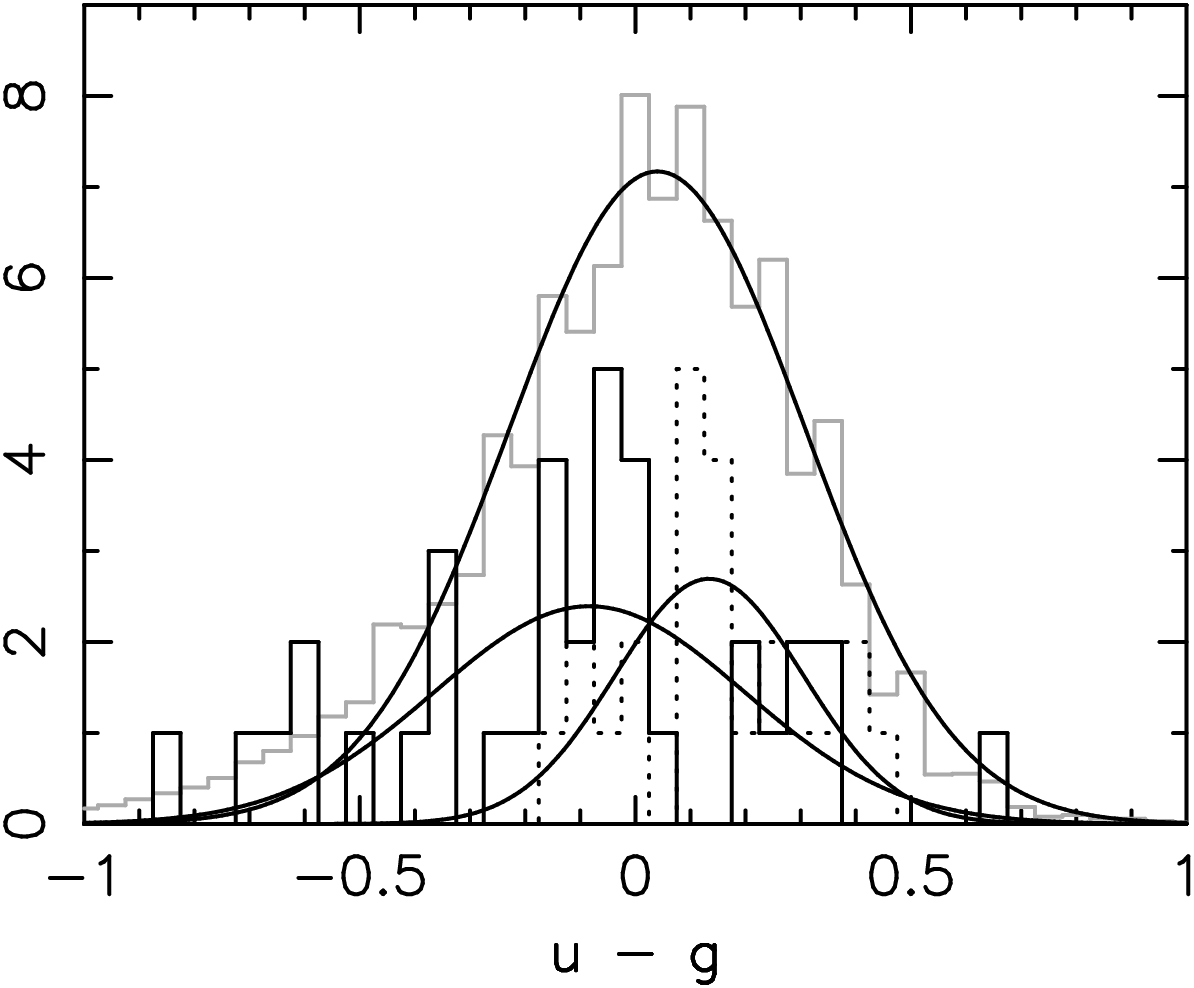}
\hspace{5pt}
\includegraphics[angle=-90,width=0.31\columnwidth]{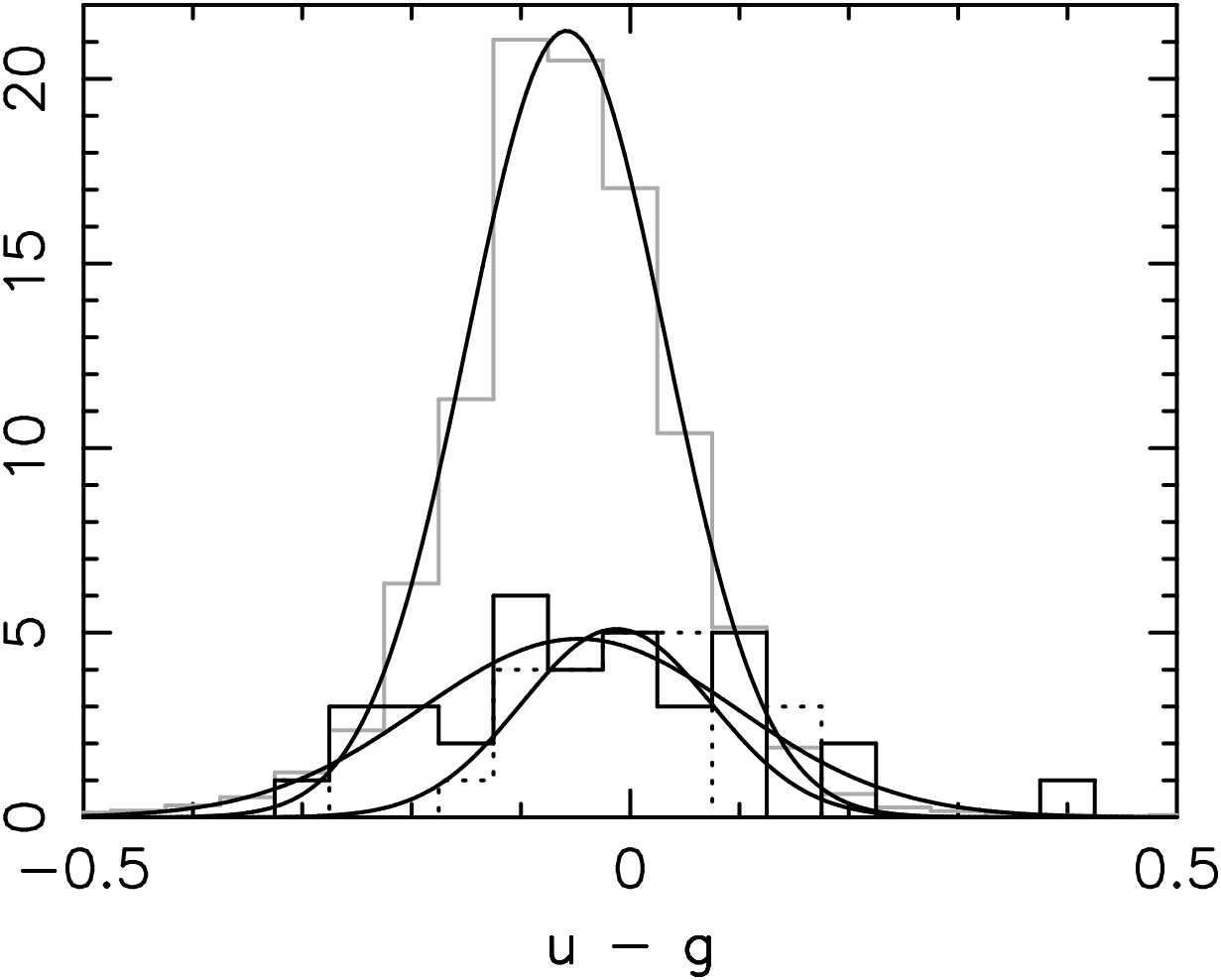}
\hspace{5pt}
\includegraphics[angle=-90,width=0.31\columnwidth]{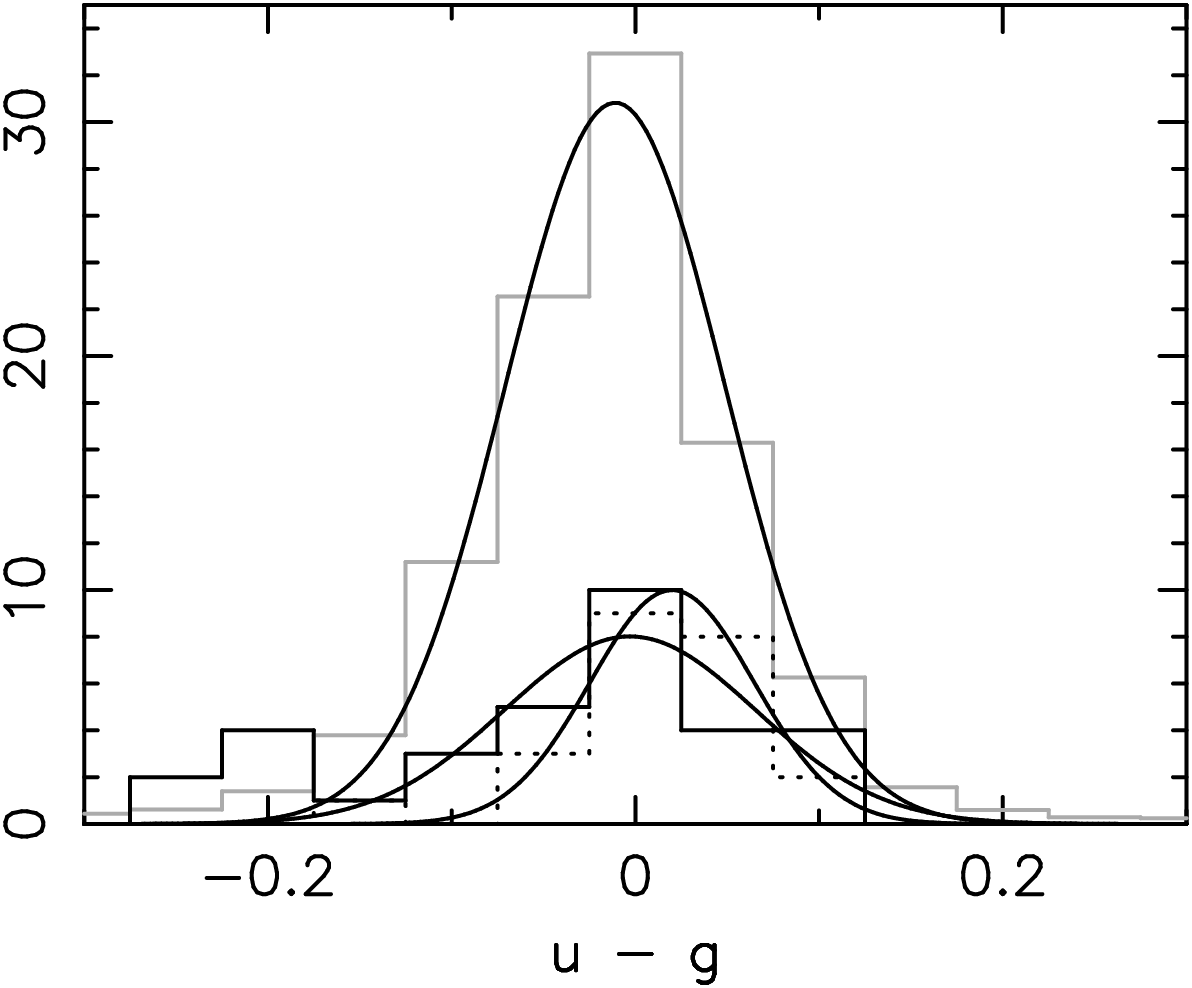}
\figcaption{
{\sl Left-to-right:} \umg, \gmr, and \rmi\ color distributions for all resolution 
elements in all galaxies (light lines), all ULX regions (heavy lines), and all control regions (dotted lines) with significance $\ge$3$\sigma$
above the local sky background. Smooth curves are Gaussian function fits to the 
histograms.
\label{f:sumcolors}}
\end{center}
\end{figure*}

A comparison of the color distributions of all 100$\times$100~pc$^2$ regions 
 in all ULX host galaxies with the color distribution of ULX regions and with 
 the color distribution of the
 control sample is shown in Figure~\ref{f:sumcolors}.
The three panels present this comparison for the three colors \umg, \gmr, and \rmi.
In each panel, the contributions from each galaxy are 
shifted along the abscissa such that regions that exactly meet
 the color criterion of Equation~1 have a color of zero. 
For clarity of presentation, 
 the histograms representing the contributions from all regions (solid histograms)
 are normalized to 100. 
Only resolution elements brighter than 3$\sigma$ above the mean sky level, including those
 containing ULXs and the control sample, are included in this census.

Also shown, in each panel, are best-fitting Gaussian functions fitted to each
 distribution. The best-fit parameter values are listed in Table~4.

The ULX \umg\ color distribution is significantly bluer than that of their host galaxies 
and of the control sample. 
The ULX color distributions in \gmr\ and \rmi\ are similar to their host galaxies
 but still bluer than the control sample. 
The ULX color distributions are broad in all 3 colors indicating the range of 
 environments in which they are found from red galaxy bulges and dust lanes to blue 
 young star forming regions. 
The trends also show, again, the strong dependence on the underlying stellar population
 to the determination of colors of individual regions.

\subsection{Qualitative considerations} \label{s:other}

We inspected our color maps and the on-line SDSS images
and consulted the literature for additional information on the
optical properties of the environments of the ULXs in our sample.
We checked whether some of the group $N$ objects may be 
 star-forming regions that are heavily reddened by dust 
 either local to the region or superimposed along the line of sight. 
Only two of the 14 group $N$ regions
 can be confidently associated with strong dust lanes based on
 the morphology of their colors. 
The X-ray luminosities of these two objects are among the 
 highest in the sample: 11.1 and 11.3$\times$10$^{39}$~\ergl.
Visual inspection shows that 5 other group $N$ regions
 are clearly within the bulges of their respective hosts. 
Six others appear to be in disk regions near to but not coincident with
 blue star clusters. 
These are designated ``patchy blue'' in Table~2 along with several
 of the group $R$ objects.
(The remaining group $N$ region appears to be a bright star cluster. 
Its colors are inconsistent with those of individual stars and of background galaxies.
It is designated ``star cluster?'' in Table~2.)
Thus, at most only a few of the ULXs in group $N$ can possibly be 
 reddened group $S$ objects.
We also note that galaxies hosting group $N$ regions are not systematically 
 stronger (dust-rich) FIR emitters than are group $S$ hosts.
Neither are they more inclined to our line-of-sight than are group $S$ hosts.
(Of course, some galaxies host both group $N$ and group $S$ objects).
 
Visual inspections also showed 
 four ULX regions with relatively strong emission in the 
 $r$ filter.  
Two of these are designated group $R$ and two are group $N$ regions
 with \gmr\ colors redder than typical of their host galaxy's \hii\ regions.
Lacking supporting evidence
 at this time, we speculate that these regions are isolated \hii\ regions whose
 $r$-band emission is dominated by \ha\ with only 
 weak contributions from underlying starlight.
These are denoted by ``\ha ?'' in the Comments column of Table~2.

\section{Discussion}

We have devised simple criteria,
 calibrated on the SDSS optical colors of known \hii\ regions in M\,81 and M\,101,
 to distinguish young, star-forming regions in nearby galaxies. 
We have used these criteria to examine the 
 100$\times$100~pc$^2$ regions around ULX candidates to determine
 if they are associated with the young stellar population.
ULX environments are designated
as ``star-forming'' if they are as blue as, or bluer
than a typical \hii\ region in that galaxy (after taking also
into account the reddening effects of dust and of old underlying
stellar populations by scaling by the mean colors of that galaxy relative to M\,81).
A useful feature of this method is that colors are distance-independent
quantities, and are measured on distance-independent regions.

For the ULX regions bright enough to permit a quantitative
analysis of their colors, we found that 60\% are bluer than
their host galaxy's \hii\ regions in all three SDSS colors.
This indicates a significant correlation between these ULXs
and young OB associations. 
However, none of them is associated
with a compact super-star-cluster within a 100$\times$100~pc$^2$ region.
This is at odds with models that invoke O-star collision and merger
processes in the collapsed core of a super-star-cluster
\citep{pz04,freitag06},
leading to the formation of an IMBH powering the ULX.
It is possible that some ULXs do indeed form this way
but this does not appear to be, by any measure, a common formation mechanism.
Conversely, there
are several ULXs located in the bulges of early-type spirals. 
We interpret them as members of a second, low-mass-donor, population
in contrast with the high-mass systems associated with recent
and on-going star formation.

Between these two extremes are several ULXs 
near patchy blue regions
of local star formation.
At the same time, we found that 10 out of the 12 most luminous
 ULXs in our sample (those with $L_{\rm X} \ga 3 \times 10^{39}$~\ergl) 
 are located in regions classified as faint or non-starforming. 
Perhaps a couple of these luminous ULX regions are heavily reddened by dust.
Note that the probability of finding a ULX outside a star-forming 
region is 0.5532 (26 of 47 ULXs are classified as either $R$ or $N$)
therefore the probability of getting 10 or more non-$S$ regions in 12 trials
selected at random from our sample is 4.4\%.

For a more physical interpretation, we recall
that our calibrator for star formation is the optical color
distribution of \hii\ regions. We speculate that the most luminous ULXs
(or equivalently, the most common phases of very high mass transfer)
are biased towards early B-type donors with an initial mass of
$\approx$10$-$15~\msun\ and an age $\sim$10$-$20~Myr
(perhaps at the stage where the B star expands to become
a blue supergiant). In that case, we expect very little residual
H$\alpha$ emission from their surroundings, the O stars having
already evolved and died. The ULX region in this scenario will appear
redder than surrounding \hii\ regions where star formation
is still ongoing.
Individual studies of stellar populations around some luminous ULXs
are consistent with this explanation: for example,
an age $\approx$10$-$20 Myr was inferred for the stars around
NGC\,4559 X-1 \citep{soria05}, and $\approx$20 Myr for those
around NGC\,1313 X-2 \citep{grise08}.

Another interpretation of the observed separation between luminous X-ray sources and 
 nearby star-forming regions \citep{zezas02, kaaret04} is that X-ray binaries
 are expelled from their place of birth.
This can occur through a variety of processes such as an asymmetric supernova
 that forms the compact remnant of the binary
 or interactions between the binary and other stars in
 the cluster \citep[see][for further discussion]{kaaret04}. 
Typical runaway velocities are of order 10 km~s$^{-1}$ and typical separations between
 X-ray binaries and star clusters are $\sim$200~pc implying ages of $\sim$20~Myrs.
However, 
\citet{kaaret04} also find the cluster-binary separation is shorter for more luminous
 X-ray sources suggesting a higher system mass (lower kick velocity) for these objects.
This is 
 inconsistent with the general pattern described here where 10 of the 12 most luminous
 ULXs are further from star-forming regions than are the majority of fainter ULXs.

The possibility that young, relatively high-mass, systems
may be located outside well-defined star-forming regions
implies there remains an observational degeneracy between
the $\la$20~Myr-old high-mass-donor and the Gyr-old low-mass-donor ULX populations.
This degeneracy is compounded by the reverse scenario; 
an old low-mass system may be located within,
or along the line-of-sight towards, a star-forming region.

However, in a more general sense, we have shown that regions of nearby galaxies 
 with differing recent star-formation histories can be separated using SDSS colors.
Hopefully, this method can be applied more generally to identify and study
correlations between classes of sources with stellar populations or 
substructures in nearby galaxies.

\acknowledgements

This work was supported, in part, by Chandra Award GO6-7081A issued by the Chandra X-ray Observatory Center which is operated by the Smithsonian Astrophysical Observatory for and on behalf of NASA under contract NAS8-03060. This work has made extensive use of DR6 of the
Sloan Digitized Sky Survey. Funding for the SDSS and SDSS-II has been provided by the Alfred P. Sloan Foundation, the Participating Institutions, the National Science Foundation, the U.S. Department of Energy, the National Aeronautics and Space Administration, the Japanese Monbukagakusho, the Max Planck Society, and the Higher Education Funding Council for England.

\bibliography{ms}

\clearpage
\begin{deluxetable}{lrrlrrrrrl}
\tablecolumns{10}
\tablewidth{0pt}
\tablenum{1} 
\tablecaption{Properties of SDSS Galaxies} 
\tablehead{
\colhead{Galaxy} & \colhead{R.A.} &  \colhead{Decl.} & \colhead{Type} & \colhead{$D$} & \colhead{$D_{25}$}  & \colhead{$u-g$} & \colhead{$g-r$} & \colhead{$r-i$} & \colhead{$f^b_{{\rm HII}}$} \\
\colhead{} &  \colhead{(h m s)} & \colhead{(d m s)} & \colhead{} & \colhead{(Mpc)}  & \colhead{(min)} & \colhead{(mag)$^a$} & \colhead{(mag)$^a$} & \colhead{(mag)$^a$} & \colhead{} 
}
\startdata
\cutinhead{SDSS Galaxies inclined $<$65\arcdeg\ hosting ULXs}
 NGC 1068 & 02 42 40.7 & $-$00 00 48 & (R)SA(rs)b & 14.4 &   7.1 & 1.24 & 0.59 & 0.31 & 0.36\\
 NGC 2500 & 08 01 53.2 & $+$50 44 14 & SB(rs)d & 10.1 &   2.9 & 0.81 & 0.30 & 0.18 & 0.29\\
 NGC 2541 & 08 14 40.1 & $+$49 03 41 & SA(s)cd & 10.6 &   6.3 & 0.59 & 0.33 & 0.13 & 0.18\\
 NGC 2681 & 08 53 32.7 & $+$51 18 49 & (R')SAB(rs)0/a & 13.3 &   3.6 & 1.44 & 0.58 & 0.35 &0.39 \\
 NGC 2841 & 09 22 02.6 & $+$50 58 35 & SA(r)b & 12.0 &   8.1 & 1.48 & 0.70 & 0.41 & 0.39\\
 NGC 3031 & 09 55 33.2 & $+$69 03 55 & SA(s)ab & 3.6 &  26.9 & 1.30 & 0.63 & 0.36 &0.38\\
 NGC 3184 & 10 18 17.0 & $+$41 25 28 & SAB(rs)cd & 8.7 &   7.4 & 1.13 & 0.47 & 0.27 &0.33\\
 NGC 3239 & 10 25 04.9 & $+$17 09 49 & IB(s)m(pec) & 8.1 &   5.0 & 0.67 & 0.23 & 0.31 &0.23\\
 NGC 3627 & 11 20 15.0 & $+$12 59 30 & SAB(s)b & 6.6 &   9.1 & 1.31 & 0.57 & 0.32 &0.38\\
 NGC 3675 & 11 26 08.6 & $+$43 35 09 & SA(s)b & 12.8 &   5.9 & 1.40 & 0.70 & 0.40 &0.33\\
 UGC 6850 & 11 52 37.3 & $-$02 28 10 & Pec & 14.5 &   0.6 & 0.47 & 0.12 & -0.13 &0.23\\
 NGC 4136 & 12 09 17.7 & $+$29 55 39 & SAB(r)c & 9.7 &   4.0 & 0.92 & 0.34 & 0.18 &0.29\\
 NGC 4150 & 12 10 33.6 & $+$30 24 06 & SA(r)0 & 9.7 &   2.3 & 1.43 & 0.61 & 0.39 &0.17\\
 NGC 4204 & 12 15 14.3 & $+$20 39 32 & SB(s)dm & 7.9 &   3.6 & 0.64 & 0.36 & 0.18 &0.31\\
 NGC 4449 & 12 28 11.9 & $+$44 05 40 & IBm & 3.0 &   6.2 & 0.88 & 0.41 & 0.16 &0.30\\
 NGC 4490 & 12 30 36.4 & $+$41 38 37 & SB(s)d(pec) & 7.8 &   6.3 & 0.87 & 0.30 & 0.13 &0.22\\
 NGC 4559 & 12 35 57.7 & $+$27 57 35 & SAB(rs)cd & 9.7 &  10.7 & 0.91 & 0.35 & 0.21 &0.26\\
 NGC 4561 & 12 36 08.1 & $+$19 19 21 & SB(rs)dm & 12.3 &   1.5 & 0.80 & 0.22 & 0.13 &0.34\\
 NGC 4618 & 12 41 32.8 & $+$41 09 03 & SB(rs)m & 7.3 &   4.2 & 0.90 & 0.37 & 0.17 &0.40\\
 NGC 4625 & 12 41 52.7 & $+$41 16 25 & SAB(rs)m(pec) & 8.2 &   2.2 & 1.06 & 0.51 & 0.20 &0.37 \\
 NGC 4725 & 12 50 26.6 & $+$25 30 03 & SAB(r)ab(pec) & 12.4 &  10.7 & 1.37 & 0.65 & 0.37 &0.31\\
 UGC 8041 & 12 55 12.6 & $+$00 07 00 & SB(s)d & 14.2 &   3.1 & 0.72 & 0.33 & 0.16 &0.19\\
 NGC 5055 & 13 15 49.3 & $+$42 01 45 & SA(rs)bc & 7.2 &  12.6 & 1.16 & 0.63 & 0.35 &0.34\\
 NGC 5204 & 13 29 36.5 & $+$58 25 07 & SA(s)m & 4.8 &   5.0 & 0.68 & 0.24 & 0.12 &0.22\\
 NGC 5457 & 14 03 12.6 & $+$54 20 57 & SAB(rs)cd & 5.4 &  28.8 & 1.04 & 0.42 & 0.23 &0.21\\
 NGC 5474 & 14 05 01.6 & $+$53 39 44 & SA(s)cd(pec) & 6.0 &   4.8 & 0.91 & 0.37 & 0.18 &0.28\\
 NGC 5585 & 14 19 48.2 & $+$56 43 45 & SAB(s)d & 7.0 &   5.7 & 0.76 & 0.27 & 0.16 &0.16\\
\cutinhead{SDSS Galaxies inclined $<$65\arcdeg\ not hosting ULXs}
 NGC 14   & 00 08 46.4 & $+$15 48 56 & (R)IB(s)m(pec) & 12.8 &   2.8 & 1.01 & 0.39 & 0.22 &0.40\\
 NGC 2537 & 08 13 14.6 & $+$45 59 23 & SB(s)m(pec) & 9.0 &   1.7 & 1.00 & 0.47 & 0.24 &0.38\\
 NGC 2903 & 09 32 10.1 & $+$21 30 03 & SAB(rs)bc & 6.3 &  12.6 & 1.18 & 0.53 & 0.33 &0.32\\
 NGC 2976 & 09 47 15.4 & $+$67 54 59 & SAc(pec) & 2.1 &   5.9 & 1.26 & 0.52 & 0.28 &0.35\\
 NGC 3077 & 10 03 19.1 & $+$68 44 02 & I0(pec) & 2.1 &   5.4 & 1.47 & 0.64 & 0.28 &0.34\\
 NGC 3274 & 10 32 17.3 & $+$27 40 08 & SABd & 5.9 &   2.1 & 0.76 & 0.27 & 0.10 &0.23\\
 NGC 3319 & 10 39 09.5 & $+$41 41 13 & SB(rs)cd & 11.5 &   6.2 & 0.66 & 0.36 & 0.22 &0.17\\
 NGC 3344 & 10 43 31.1 & $+$24 55 20 & (R)SAB(r)bc & 6.1 &   7.1 & 1.01 & 0.42 & 0.30 &0.23\\
 NGC 3351 & 10 43 57.7 & $+$11 42 13 & SB(r)b & 8.1 &   7.4 & 1.28 & 0.61 & 0.30 &0.37\\
 NGC 3368 & 10 46 45.7 & $+$11 49 12 & SAB(rs)ab & 8.1 &   7.6 & 1.35 & 0.64 & 0.34 &0.33\\
 NGC 3486 & 11 00 23.9 & $+$28 58 29 & SAB(r)c & 7.4 &   7.1 & 0.82 & 0.34 & 0.22 &0.14\\
 NGC 3738 & 11 35 48.8 & $+$54 31 26 & Im & 4.3 &   2.5 & 1.13 & 0.41 & 0.18 &0.27\\
 NGC 3985 & 11 56 42.1 & $+$48 20 02 & SB(s)m & 8.3 &   1.3 & 1.00 & 0.45 & 0.25 &0.47\\
 NGC 4020 & 11 58 56.9 & $+$30 24 49 & SBd & 8.0 &   2.1 & 0.93 & 0.40 & 0.21 &0.31\\
 NGC 4062 & 12 04 03.8 & $+$31 53 45 & SA(s)c & 9.7 &   4.1 & 1.25 & 0.58 & 0.33 &0.33\\
 NGC 4203 & 12 15 05.0 & $+$33 11 50 & SAB0- & 9.7 &   3.4 & 1.62 & 0.73 & 0.39 &0.28\\
 NGC 4207 & 12 15 30.4 & $+$09 35 06 & S & 8.3 &   1.6 & 1.36 & 0.64 & 0.33 &0.37\\
 NGC 4214 & 12 15 39.2 & $+$36 19 37 & IAB(s)m & 3.5 &   8.5 & 0.88 & 0.34 & 0.20 &0.23\\
 NGC 4245 & 12 17 36.8 & $+$29 36 29 & SB(r)0/a & 9.7 &   2.9 & 1.53 & 0.68 & 0.38 &0.36\\
 NGC 4309 & 12 22 12.4 & $+$07 08 39 & SAB(r)0+ & 11.9 &   1.8 & 1.37 & 0.68 & 0.35 &0.31\\
 NGC 4310 & 12 22 26.3 & $+$29 12 31 & (R')SAB(r)0+ & 9.7 &   2.2 & 1.33 & 0.68 & 0.41 &0.34\\
 NGC 4314 & 12 22 32.0 & $+$29 53 43 & SB(rs)a & 9.7 &   4.2 & 1.55 & 0.71 & 0.38 &0.28\\
 NGC 4370 & 12 24 54.9 & $+$07 26 40 & Sa & 10.7 &   1.4 & 1.62 & 0.74 & 0.39 &0.40\\
 NGC 4395 & 12 25 48.9 & $+$33 32 48 & SA(s)m & 3.6 &  13.2 & 0.78 & 0.27 & 0.16 &0.26\\
 NGC 4414 & 12 26 27.1 & $+$31 13 25 & SA(rs)c & 9.7 &   3.6 & 1.24 & 0.58 & 0.33 &0.39\\
 NGC 4491 & 12 30 57.1 & $+$11 29 01 & SB(s)a & 6.8 &   1.7 & 1.41 & 0.65 & 0.34 &0.37\\
 NGC 4509 & 12 33 06.8 & $+$32 05 30 & Sab(pec) & 12.8 &   0.9 & 0.65 & 0.21 & 0.05 &0.33\\
 IC 3521  & 12 34 39.5 & $+$07 09 37 & IBm & 8.2 &   1.0 & 1.20 & 0.62 & 0.30 &0.39\\ 
 NGC 4627 & 12 41 59.7 & $+$32 34 25 & E4(pec) & 7.4 &   2.6 & 1.25 & 0.42 & 0.22 &0.33\\
 NGC 4670 & 12 45 17.1 & $+$27 07 32 & SB(s)0/a(pec) & 11.0 &   1.4 & 0.70 & 0.42 & 0.19 &0.17\\
 NGC 5949 & 15 28 00.7 & $+$64 45 47 & SA(r)bc & 11.2 &   2.2 & 1.12 & 0.57 & 0.28 &0.41\\
\enddata

\tablenotetext{a}{Average SDSS colors within \DTF\ area of galaxy.}
\tablenotetext{b}{Fractional area of galaxy blueward of \hii\ regions according to Equation~1.}

\end{deluxetable}

\clearpage

\begin{deluxetable}{rrlrrrrrll}
\tablecolumns{10}
\tablewidth{0pt}
\tablenum{2} 
\tablecaption{Properties of ULX Candidates} 
\tablehead{
\colhead{R.A.} &  \colhead{Decl.} & \colhead{Galaxy} & \colhead{$L_{\rm X}^a/10^{39}$}
& \colhead{$u-g^b$} & \colhead{$g-r^b$} & \colhead{$r-i^b$} & \colhead{$\sigma_{\rm min}^c$}
& \colhead{Group$^d$} & \colhead{Comments} \\
\colhead{(h m s)} & \colhead{(d m s)} & \colhead{Host} & \colhead{(\ergl)}
& \colhead{(mag)} & \colhead{(mag)} & \colhead{(mag)} & \colhead{} & \colhead{} & \colhead{}
}
\startdata
02 42 38.89 & -00 00 55.1 & NGC1068 & 8.0(11.3) & 1.18 & 0.73 & 0.31 & 11.0& N & dust lane\\
02 42 39.71 & -00 01 01.4 & NGC1068 & 3.0(1.4) & 1.07 & 0.41 & 0.26 & 28.2 & S & \\
02 42 40.43 & -00 00 52.6 & NGC1068 & 1.3(0.6) & 1.60 & 0.75 & 0.32 & 47.2 & N & bulge\\
08 01 48.10 & 50 43 54.6 & NGC2500 & 6.0(c2f) & 0.88 & 0.24 & 0.04 & 4.0   & N(S) & patchy blue\\
08 01 57.85 & 50 43 39.5 & NGC2500 & 1.2(c2f) & 0.58 & 0.16 & 0.26 & 14.5  & N & star cluster? \\
08 14 37.02 & 49 03 26.6 & NGC2541 & 4.4(c2f) & 0.74 & 0.24 & 0.25 & 1.3   & R & patchy blue\\
08 53 33.66 & 51 19 29.5 & NGC2681 & 1.2(0.9) & 1.34 & 0.59 & 0.38 & 2.2   & R & \\
08 53 35.75 & 51 19 17.3 & NGC2681 & 0.7(1.8) & 1.06 & 0.65 & 0.38 & 3.3   & S & \\
09 22 02.22 & 50 58 54.2 & NGC2841 & 1.1(0.7) & 1.81 & 0.87 & 0.47 & 13.3    & N & bulge\\
09 55 32.97 & 69 00 33.4 & NGC3031 & 2.9(3.8) & 1.20 & 0.69 & 0.35 & 13.2  & S(N) &\\
10 18 12.05 & 41 24 20.7 & NGC3184 & 1.0(1.7) & 1.88 & 0.45 & 0.30 & 2.2   & R & patchy blue\\
10 18 23.00 & 41 27 41.7 & NGC3184 & 1.1(2.1) & 0.51 & 0.29 & 0.06 & 3.8   & S &\\
10 25 06.98 & 17 09 47.2 & NGC3239 & 1.4(c2f) & 0.66 & 0.34 & 0.14 & 13.9  & S & \\
10 25 08.20 & 17 09 48.3 & NGC3239 & 1.9(c2f) & 0.55 & 0.36 & 0.06 & 8.1   & S &\\
11 20 18.31 & 12 59 00.3 & NGC3627 & 1.2(c2f) & 0.98 & 0.50 & 0.27 & 13.7  & S & \\
11 20 20.89 & 12 58 46.1 & NGC3627 & 4.1(c2f) & 1.77 & 0.53 & 0.32 & 1.4   & R &\\
11 26 07.33 & 43 34 06.3 & NGC3675 & 1.7(c2f) & 1.73 & 0.71 & 0.44 & 3.0   & N & patchy blue\\
11 52 37.36 & 02 28 07.1 & UGC6850 & 1.8(c2f) & 0.41 & -0.06 & -0.34 & 13.0& S & \\
12 09 22.18 & 29 55 59.7 & NGC4136 & 2.1(1.7) & 0.35 & 0.26 & -0.07 & 4.5  & S & \\
12 10 33.76 & 30 23 58.0 & NGC4150 & 3.9(c2f) & 1.61 & 0.75 & 0.39 & 17.7  & N & bulge\\
12 15 10.91 & 20 39 12.4 & NGC4204 & 1.6(c2f) & -0.06 & 0.33 & -0.44 & 1.8 & R & isolated \hii ?\\
12 28 17.83 & 44 06 33.9 & NGC4449 & 0.6(0.8) & 0.17 & 0.25 & -0.19 & 63.7 & S &\\
12 30 29.55 & 41 39 27.6 & NGC4490 & 1.1(1.0) & 0.81 & 0.75 & -0.30 & 15.5 & N(S) & patchy blue, \hii ?\\
12 30 30.82 & 41 39 11.5 & NGC4490 & 2.7(2.6) & 0.54 & 0.30 & 0.18 & 18.7  & N & patchy blue\\
12 30 32.27 & 41 39 18.1 & NGC4490 & 2.0(1.7) & 0.74 & 0.40 & 0.10 & 14.1  & S & \\
12 30 36.32 & 41 38 37.8 & NGC4490 & 2.1(1.8) & 1.12 & 0.37 & 0.22 & 45.3  & N & bulge\\
12 30 43.26 & 41 38 18.4 & NGC4490 & 3.8(3.1) & 0.68 & 0.26 & 0.10 & 14.8  & S & \\
12 35 51.71 & 27 56 04.1 & NGC4559 & 22.9(9.4) & 0.28 & -0.08 & -0.06 & 1.8& R & patchy blue \\
12 35 57.79 & 27 58 07.4 & NGC4559 & 2.8(1.8) & 0.76 & 0.32 & 0.22 & 21.4  & S(N) & \\
12 35 58.56 & 27 57 41.9 & NGC4559 & 15.0(11.1) & 1.11 & 0.59 & 0.31 & 15.5& N & dust lane \\
12 36 08.90 & 19 19 55.9 & NGC4561 & 3.0(c2f) & 0.57 & 0.35 & -0.04 & 1.9  & R & patchy blue \\
12 41 29.14 & 41 07 57.7 & NGC4618 & 1.7(c2f) & 0.54 & 0.23 & 0.09 & 4.6   & S & \\
12 41 52.72 & 41 16 31.7 & NGC4625 & 1.3(c2f) & 1.00 & 0.51 & 0.21 & 20.5  & S(N) & \\
12 50 25.70 & 25 31 29.8 & NGC4725 & 1.6(1.8) & 0.50 & 0.60 & -0.10 & 9.1    & S & \hii ? \\
12 50 26.37 & 25 33 19.4 & NGC4725 & 2.1(2.0) & -0.21 & 0.41 & 1.11 & 0.7    & R & \\
12 50 27.39 & 25 30 26.5 & NGC4725 & 1.0(0.8) & 2.04 & 0.88 & 0.48 & 6.4     & N & bulge \\
12 50 36.88 & 25 30 28.4 & NGC4725 & 1.8(1.6) & 1.23 & 0.79 & 0.52 & 0.9     & R & \\
12 55 12.31 & 00 07 51.9 & UGC8041 & 2.4(c2f) & 0.24 & 0.08 & 0.05 & 6.0   & S & \\
13 15 19.54 & 42 03 02.3 & NGC5055 & 6.0(13.5) & 0.73 & 0.40 & 0.30 & 2.2  & R & \\
13 15 39.33 & 42 01 53.4 & NGC5055 & 0.4(0.6) & 1.17 & 0.69 & 0.34 & 8.2   & S(N) & \\
13 16 02.27 & 42 01 53.6 & NGC5055 & 1.5(1.3) & 1.12 & 0.69 & 0.18 & 6.5   & S & \\
13 29 38.61 & 58 25 05.6 & NGC5204 & 4.7(1.2) & 0.69 & 0.23 & 0.14 & 26.8  & S(N) & \\
14 03 32.39 & 54 21 02.9 & NGC5457 & 3.4(2.0) & 0.96 & 0.34 & 0.23 & 8.6   & S(N) & \\
14 04 14.29 & 54 26 03.8 & NGC5457 & 1.7(1.2) & 0.52 & 0.08 & 0.74 & 1.2   & R & \\
14 04 59.74 & 53 38 08.9 & NGC5474 & 5.1(c2f) & 0.92 & 0.30 & 0.25 & 8.4   & N & patchy blue \\
14 19 39.39 & 56 41 37.8 & NGC5585 & 2.8(c2f) & -0.31 & 0.78 & -0.60 & 1.1 & R & isolated \hii ? \\
14 19 52.14 & 56 44 17.3 & NGC5585 & 1.0(c2f) & 1.08 & 0.38 & 0.28 & 3.7   & N &  patchy blue\\
\enddata

\tablenotetext{a}{Observed luminosity estimated from count rates, values from spectral fits in parentheses. ``c2f'' denotes too few counts for spectral fitting.}
\tablenotetext{b}{Average colors within 100$\times$100 pc$^2$ region.}
\tablenotetext{c}{SDSS detection significance above sky level in $\sigma$.}
\tablenotetext{d}{Group designation $R$ for rejected, $S$ for star-forming,
 and $N$ for non-star-forming according to the empirical criteria, \S\ref{s:ECC}; values in parentheses according to the Fisher criterion, \S\ref{s:LDA}, when different.}

\end{deluxetable}

\clearpage

\begin{center}
\small{
\begin{tabular}{lccccc}
\multicolumn{6}{c}{{\sc TABLE 3}} \\
\multicolumn{6}{c}{{\sc SDSS Color Distributions}} \\
\hline \hline
M\,81 & \umg & \gmr & \rmi & $f_{Empirical}$ & $f_{Fisher}$ \\
\hline
Galaxy ($\langle C_{j,\rm M81} \rangle$)& 1.30$\pm$0.34 & 0.63$\pm$0.11 & 0.36$\pm$0.09 & &\\
\hii\  ($h_j\pm\delta h_j$) & 1.03$\pm$0.30 & 0.61$\pm$0.10 & 0.31$\pm$0.09 & 77.0 & 64.4\\
Disk   & 1.26$\pm$0.26 & 0.62$\pm$0.08 & 0.36$\pm$0.09 & &\\
Bulge  & 1.66$\pm$0.10 & 0.74$\pm$0.04 & 0.41$\pm$0.03 & &\\
\hline
M\,101 & \umg & \gmr & \rmi & $f_{Empirical}$ & $f_{Fisher}$ \\
\hline
Galaxy & 1.04$\pm$0.24 & 0.42$\pm$0.10 & 0.23$\pm$0.06 & &\\
\hii\  & 0.93$\pm$0.26 & 0.42$\pm$0.13 & 0.17$\pm$0.10 & 68.6 & 69.5 \\
\hline
\multicolumn{6}{l}{$f_{Empirical}$ and $f_{Fisher}$ are the fractions of regions classified as bluer than \hii\ regions}\\
\multicolumn{6}{l}{according to the criteria of \S\ref{s:ECC} and \S\ref{s:LDA}, respectively} \\
\end{tabular}
} 
\end{center}

\begin{center}
\small{
\begin{tabular}{lrrrrrr}
\multicolumn{7}{c}{{\sc TABLE 4}} \\
\multicolumn{7}{c}{{\sc Gaussian Fits to Color Distributions}} \\
\hline \hline
& \multicolumn{2}{c}{Galaxies} & \multicolumn{2}{c}{ULXs} & \multicolumn{2}{c}{Control} \\ 
     & \multicolumn{1}{c}{Center} & \multicolumn{1}{c}{$\sigma$} & \multicolumn{1}{c}{Center} & \multicolumn{1}{c}{$\sigma$} & \multicolumn{1}{c}{Center} & \multicolumn{1}{c}{$\sigma$} \\ \hline
\umg &  0.040 & 0.268 & -0.084 & 0.284 &  0.135 & 0.172\\
\gmr & -0.058 & 0.091 & -0.047 & 0.145 & -0.013 & 0.087 \\
\rmi & -0.011 & 0.060 & -0.003 & 0.069 &  0.020 & 0.045 \\
\hline
\end{tabular}
} 
\end{center}

\clearpage

\newpage

\end{document}